\documentclass{article}

\usepackage[final]{neurips_2025}

\usepackage[utf8]{inputenc} 
\usepackage[T1]{fontenc}    
\usepackage{hyperref}       
\usepackage{url}            
\usepackage{booktabs}       
\usepackage{amsfonts}       
\usepackage{nicefrac}       
\usepackage{microtype}      
\usepackage{xcolor}         
\usepackage{multirow}
\usepackage{amsmath}
\usepackage{amssymb}
\usepackage{mathtools}
\usepackage{float}
\usepackage{amsthm}
\usepackage{makecell}
\usepackage{wrapfig}
\usepackage{graphicx}
\usepackage{subcaption}
\usepackage{threeparttable}
\usepackage{rotating}
\usepackage[textsize=tiny]{todonotes}
\usepackage[table]{xcolor}
\definecolor{simsortcolor}{rgb}{0.435,0.176,0.761} 

\author{
    Yimu Zhang \\ 
    Fudan University \\
    \texttt{email2} \\
    \And
    Dongqi Han \\
    Microsoft Research Asia \\
    \texttt{dongqihan@microsoft.com} \\
    \And
    Yansen Wang \\
    Microsoft Research Asia \\
    \texttt{email2} \\
    \And
    Zhenning Lv \\
    Fudan University \\
    \texttt{email2} \\
    \And
    Yu Gu \\
    Fudan University \\
    \texttt{guyu\_@fudan.edu.cn} \\
    \And
    Dongsheng Li \\
    Microsoft Research Asia \\
    \texttt{email2} \\
}

\title{SimSort: A Data-Driven Framework for Spike Sorting \\ by Large-Scale Electrophysiology Simulation}

\author{%
\textbf{Yimu Zhang}$^{1}$\footnotemark[1] \quad
\textbf{Dongqi Han}$^{2}$\footnotemark[2] \quad
\textbf{Yansen Wang}$^{2}$ \quad
\textbf{Zhenning Lv}$^{1}$ \\
\textbf{Yu Gu}$^{1}$\footnotemark[2] \quad
\textbf{Dongsheng Li}$^{2}$ \\
$^1$ Institutes of Brain Science, Fudan University \quad \\
$^2$Microsoft Research Asia \\
\texttt{\{yimuzhang21, znlv23\}@m.fudan.edu.cn},
\texttt{\{guyu\_\}@fudan.edu.cn},\\
\texttt{\{dongqihan,yansenwang,dongsli\}@microsoft.com} \\
}

\begin{document}

\maketitle

\footnotetext[1]{The work was conducted during the internship of Yimu Zhang at Microsoft Research Asia.}
\footnotetext[2]{Corresponding authors.}

\begin{abstract}
Spike sorting is an essential process in neural recording, which identifies and separates electrical signals from individual neurons recorded by electrodes in the brain, enabling researchers to study how specific neurons communicate and process information. 
Although there exist a number of spike sorting methods which have contributed to significant neuroscientific breakthroughs, many are heuristically designed, making it challenging to verify their correctness due to the difficulty of obtaining ground truth labels from real-world neural recordings.
In this work, we explore a data-driven, deep learning-based approach. We begin by creating a large-scale dataset through electrophysiology simulations using biologically realistic computational models. We then present \textbf{SimSort}, a pretraining framework for spike sorting. 
Trained solely on simulated data, SimSort demonstrates zero-shot generalizability to real-world spike sorting tasks, yielding consistent improvements over existing methods across multiple benchmarks.
These results highlight the potential of simulation-driven pretraining to enhance the robustness and scalability of spike sorting in experimental neuroscience.


\end{abstract}

\section{Introduction}
Understanding the complex computations performed by the brain requires insight into the activity of individual neurons \cite{lewickiReviewMethodsSpike1998, buzsakiLargescaleRecordingNeuronal2004}, which is crucial for exploring how information is encoded, processed, and transmitted within neural circuits, as well as decoding the brain's functional dynamics \cite{quianquirogaExtractingInformationNeuronal2009, joshiDynamicSynchronizationHippocampal2023, sarkarAdvancedSpikeSorting2024}. Recent advances in neural recording technologies have enabled capturing the activity of neurons across multiple regions of the brain with high spatial and temporal precision \cite{chungFullyAutomatedApproach2017, steinmetzNeuropixels20Miniaturized2021a, chungHighDensityLongLastingMultiregion2019, hongNovelElectrodeTechnologies2019}. Notably, extracting meaningful information from recordings relies on a critical step known as \textbf{spike sorting}.

\begin{figure*}[t]
    \centering
    \vspace{-3mm}
    \includegraphics[width=\linewidth]{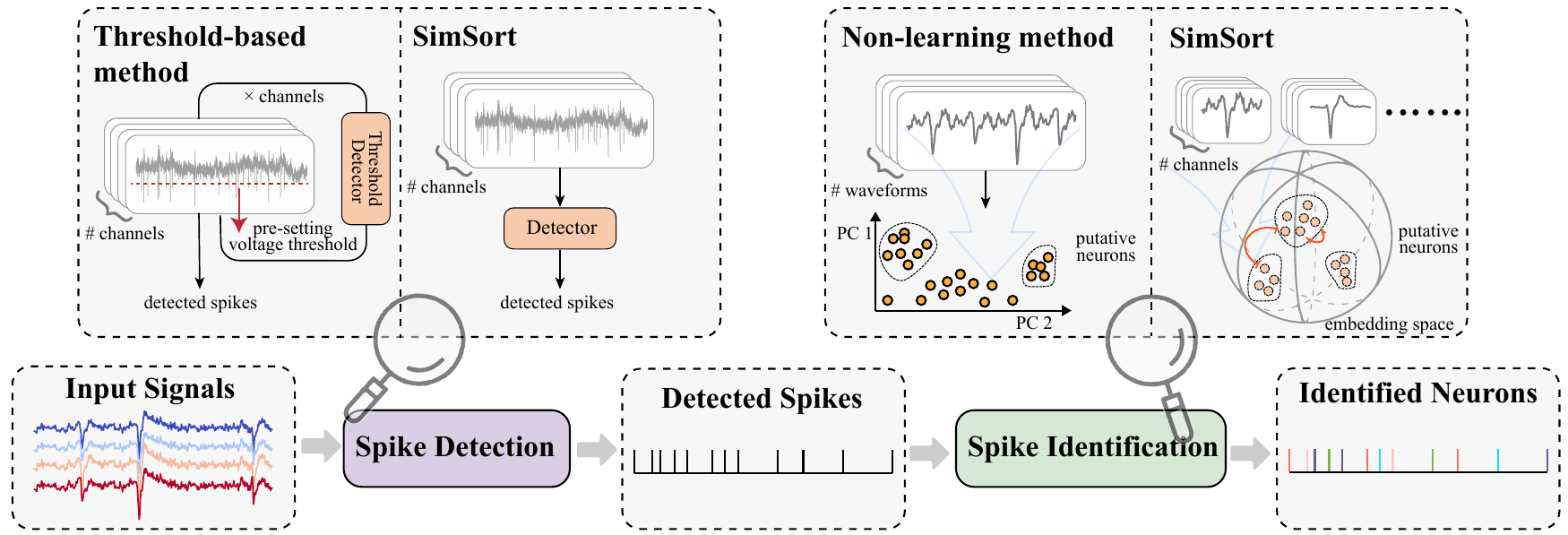}
    \caption{
    Pipeline of \textbf{spike sorting}, which consists of two main steps:\textbf{ spike detection} and \textbf{spike identification}. For spike detection, typical spike sorting algorithms (left) utilize threshold-based detector relies on fixed voltage thresholds on each channel, and use non-learning method like performing PCA-based clustering on concatenated waveforms for spike identification. Those approaches are sensitive to noise and require manual parameter tuning. In contrast, our proposed framework SimSort (right) use a neural network-based detector replaces the threshold method for spike detection, enhancing robustness and generalization. For spike identification, the feature embeddings of multi-channel waveforms learned from contrastive learning to improve clustering accuracy.
    }
    \label{fig:pipeline}
    \vspace{-3mm}
\end{figure*}

Spike sorting is the process of extracting and identifying neural activity from extracellular recordings. It involves two main steps (Fig.~\ref{fig:pipeline}): \textbf{spike detection}, which extracts spike events from background noise, and \textbf{spike identification}, which assigns these detected spikes to individual neurons \cite{lefebvreRecentProgressMultielectrode2016}. Spike sorting is indispensable for transforming raw electrical signals into interpretable data that reveal the firing patterns of individual neurons. Accurate spike sorting is essential for linking neural activity to behavior, understanding the functional organization of neural circuits \cite{sibilleHighdensityElectrodeRecordings2022}, and uncovering mechanisms underlying various sensory and cognitive processes \cite{liuDecodingCognitionSpontaneous2022}. Furthermore, its role extends to translational applications, such as improving neural prosthetics and developing closed-loop BCIs, where spike sorting is critical for achieving precise neural decoding and control \cite{park128ChannelFPGABasedRealTime2017, hanLiveDemonstrationEfficient2024}.

For a long time, spike sorting predominantly relied on heuristic statistics and machine learning approaches \cite{lewickiReviewMethodsSpike1998, quirogaUnsupervisedSpikeDetection2004, rossantSpikeSortingLarge2016, pachitariuFastAccurateSpike2016, chungFullyAutomatedApproach2017, hilgenUnsupervisedSpikeSorting2017, pachitariuSpikeSortingKilosort42024}. While these methods have promoted neuroscience research, they exhibit several key limitations. First, their sorting results are sensitive to parameter settings and post-processing, which depend on the experimenter's expertise and must be customized for each dataset. Moreover, these approaches lack a data-driven foundation, limiting their scalability and adaptability across diverse experimental settings, particularly in low signal-to-noise ratio (SNR) and high-variability scenarios, where their performance declines significantly. Recently, deep learning-based methods, such as YASS \cite{leeYASSAnotherSpike2017} and CEED \cite{vishnubhotlaRobustGeneralizableRepresentations2023}, have attempted to improve spike sorting by adopting data-driven approaches to enhance detection and clustering accuracy. However, their generalizability and practical use are hindered by the limited training datasets.

To address these limitations, we highlight the importance of using massive training data with ground-truth annotations to achieve more reliable and robust spike sorting. Furthermore, we aim to design a deep learning framework optimized for spike sorting and pretrain models for practical usage.

Therefore, in this work, we first generated a large-scale labeled dataset to address the scarcity of ground-truth data for spike sorting. Building on this, we present \textbf{SimSort}, a framework that utilizes data-driven approaches to enable fully automated spike sorting. By pretraining a spike detection model on the large-scale dataset, SimSort achieved notable improvements in detection accuracy and adaptability compared to commonly used threshold-based methods. Additionally, SimSort incorporated contrastive learning to enhance waveform feature representations, achieving better robustness against noise. We evaluated SimSort in zero-shot settings on publicly available datasets without fine-tuning, showing its effectiveness in spike sorting tasks compared to existing methods. 

The key contributions of the SimSort framework include: 

\textbf{1)} A publicly available, large-scale labeled dataset to help address the scarcity of ground-truth data, supporting the development and evaluation of learning-based spike sorting methods. 

\textbf{2)} A pretraining paradigm for spike sorting with a large-scale simulation dataset. The results, for the first time, demonstrate successful zero-shot transfer from simulated to real-world spike sorting tasks. 

\textbf{3)} An off-the-shelf pretrained model for neuroscientists to use\footnote{See \url{https://SimSortTool.github.io} for the model, code and usage instructions.}, enabling fully automated spike sorting on Tetrode recordings without dependence on hand-tuned parameters.

\vspace{-2mm}
\section{Preliminaries}  
The goal of \textbf{spike sorting} (Figure~\ref{fig:pipeline}) is to extract single-neuron spiking activity from extracellular recordings, where the activities of an unknown number of neurons are mixed together. Mathematically, spike sorting is a blind source separation problem \cite{buccinoSpikeSortingNew2022} and can be formally defined as follows: Let \( \mathbf{V} \in \mathbb{R}^{T \times C} \) represent the extracellular recording (voltage) across \( C \) electrode channels over \( T \) time points. The goal is to infer a set of spike times \( \{t_k\}_{k=1}^K \) and corresponding neuron labels \( \{y_k\}_{k=1}^K \), where \( y_k \in \{1, \dots, N\} \) represents the neuron identity, and \( N \) is the total number of detected neurons.

\section{Related Work}
\vspace{-2mm}
Early spike sorting pipelines typically used thresholds to detect spike events, followed by dimensionality reduction (e.g., PCA or $t$-SNE) and clustering \cite{FRS1901LIIIOL, maatenVisualizingDataUsing2008, veerabhadrappaCompatibilityEvaluationClustering2020}. Although this pipeline was straightforward, it often suffered from noise susceptibility, inaccurate detection of low SNR events, and reliance on manual parameter tuning.

Deep learning approaches have been adopted to replace or enhance certain steps in traditional pipelines, aiming to improve robustness, automation, and adaptability to diverse experimental conditions. Autoencoders have been explored as a method for dimensionality reduction \cite{baldiAutoencodersUnsupervisedLearning2012, wuLearningSortFewshot2019, eomDeeplearnedSpikeRepresentations2021}. YASS \cite{leeYASSAnotherSpike2017} employed a convolutional neural network for spike detection and waveform cleaning, thereby mitigating clustering errors caused by distorted waveforms. CEED \cite{vishnubhotlaRobustGeneralizableRepresentations2023} applied a contrastive learning framework to enforce invariances to amplitude fluctuations, noise, and channel subset changes in the extracted embeddings. However, these methods were limited by their reliance on restricted training datasets, potentially reducing their generalization across diverse experimental conditions.

In particular, YASS depended on high-quality prior training data and assumed a consistent experimental setup with validated sorting results, limiting its utility in scenarios where training data were scarce or recording conditions varied significantly. 

Similarly, CEED had its limitations: (1) its invariance assumptions may not generalize to datasets with unaccounted variability; (2) it relied on KiloSort2.5-processed data, potentially inheriting inaccuracies from these analyses; (3) its training and testing datasets were narrowly scoped, originating from similar experimental conditions, which may constrain broader applicability; and (4) it did not optimize spike detection or provide a complete spike sorting pipeline.

To overcome these limitations, we propose a data-driven approach to develop a fully automated spike sorting pipeline that does not rely on manually defined parameters, demonstrating improved generalization across diverse datasets and recording conditions.
\vspace{-2mm}
\section{Methods}
\vspace{-2mm}
\subsection{Dataset Creation}
\label{subsec: Data Generation}
To address the scarcity of labeled spike data, we generated a large-scale synthetic dataset using biophysically detailed neuron models from the Blue Brain Project (BBP) \cite{hayModelsNeocorticalLayer2011,markramReconstructionSimulationNeocortical2015,laquitaineSpikeSortingBiases2025}, covering 206 neuron models from layers 1–6 of the juvenile rat somatosensory cortex. These models capture diverse morphologies and electrophysiological dynamics across 30 neuronal types.

\paragraph{Electrophysiology Simulation}
Based on these models, we first simulated intracellular activity by injecting noise currents into multi-compartment neurons using the NEURON simulator \cite{hinesNEURONSimulationEnvironment1997}, producing realistic spiking responses. Then, extracellular potentials were computed via volume conductor theory, summing transmembrane currents across compartments to obtain virtual recordings from tetrode electrodes randomly placed near the neurons. Each simulation trial included 5 neurons of varying types, placed in a $100 \times 100 \times 100~\mu$m$^3$ volume, with randomized positions for both cells and electrodes. (Figure~\ref{fig:simulation}; full technical details in Appendix~\ref{appendix: A}).



\begin{figure*}[t]
    \centering
    \includegraphics[width=\linewidth]{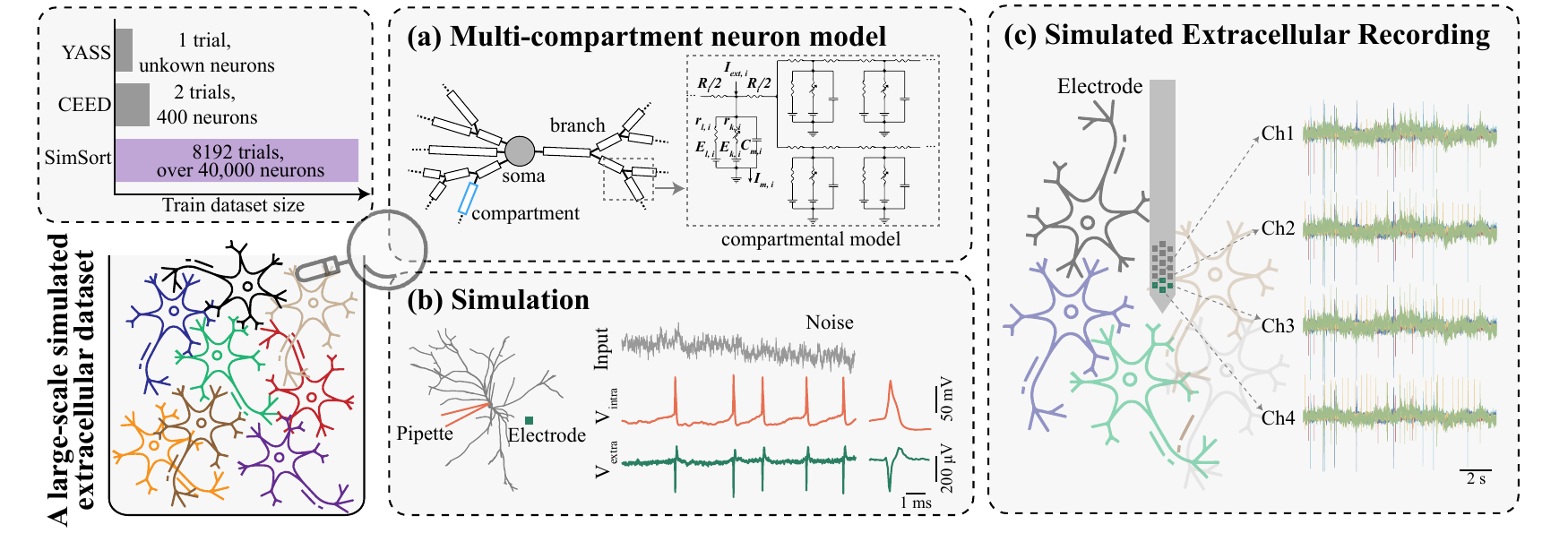}
    \caption{
        Overview of the large-scale simulated extracellular dataset generation process.
        (a) Multi-compartment neuron models simulate detailed neuronal morphologies and electrophysiological properties, incorporating realistic ion channel dynamics. 
        (b) Noise generated by stochastic process is injected into the somatic compartment to induce stochastic intracellular action potential firing, and extracellular signals are recorded using virtual electrodes placed near the neurons in a simulated environment. 
        (c) The resulting multi-channel extracellular recordings capture diverse and realistic neural activity.
    }
    \label{fig:simulation}
    \vspace{-5mm}
\end{figure*}

\paragraph{Dataset Preparation}
\label{subsec: Dataset Preparation}
We generated a large-scale dataset consisting of 8192 simulated recording trials, representing continuous neuronal activities from more than 40,000 individual neurons. Intracellular spike timestamps were recorded as ground-truth for extracellular spikes, providing precise annotations for model training and evaluation.

The dataset was structured into two primary subsets: 

\textbf{Continuous Signal Dataset}: This subset comprised full-length extracellular recordings across all electrode channels. It was designed for training spike detection models and evaluating the overall performance of spike sorting pipelines.

\textbf{Spike Waveform Dataset}: This subset contained spike waveforms extracted from each ground-truth units. It was used for training and evaluating the spike identification model.

We used signals simulated from BBP layers 1–5 neuron models for training and validation, reserving signals from BBP layer 6 neuron models exclusively for evaluation, ensured that the test set comprises previously unseen neuron types and configurations, allowing us to effectively assess the model's generalization capabilities.
\vspace{-2mm}
\subsection{Spike Detection}
\label{subsec: method-Spike detection}
We formulate the spike detection task as identifying the temporal segments $\{t_k\}$ corresponding to putative neural spikes from the raw input $\mathbf{V} \in \mathbb{R}^{T \times C}$, where $T$ represents the number of time points and $C$ denotes the number of electrode channels. The raw signal $\textbf{V}$ first underwent two preprocessing steps: bandpass filtering and spatial whitening. Then, the processed signal was subsequently provided as input to the spike detection model.

\begin{wrapfigure}{r}{0.6\textwidth}
    \vspace{-7mm}
  \begin{center}
    \includegraphics[width=1.0\linewidth]{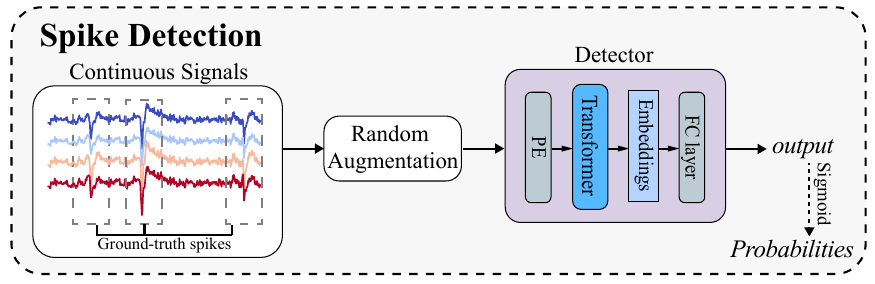}
  \end{center}
    \caption{Spike detection model in SimSort.}
    \label{fig:detection_model}
    \vspace{-2mm}
\end{wrapfigure}

We utilized a Transformer-based architecture \cite{vaswaniAttentionAllYou2017} for spike detection, trained on the simulated continuous signal dataset detailed in Sec.~\ref{subsec: Dataset Preparation}. For robustness and generalization, data augmentation was integrated into the training process, referring to Fig.~\ref{fig:detection_model}. We employed a binary cross-entropy loss for the spike detection model:
\begin{equation}
\mathcal{L}_{\text{BCE}} = -\frac{1}{T}\sum_{t=1}^T \Big(w_p\cdot  y_t \log\hat{y}_t +  (1-y_t)\log(1-\hat{y}_t)  \Big),
\end{equation}
where the binary label $y_t$ denoted spike events, \(\hat{y}_t\) was the predicted probability, \(T\) was trial length, and $w_p$ was a hyperparameter to reweigh the positive predictions given the general scarcity of spikes.

\textbf{Data Augmentation:}
\label{subsec: Data Augmentation}
To improve robustness and generalization in spike detection, data augmentation was applied to each input signal segment $V' \in \mathbb{R}^{T \times C}$ (Fig.~\ref{fig:detection_model}). A subset of channels $C' \subseteq \{1, 2, \dots, C\}$ was randomly selected with a probability $p$ to apply augmentation. Various augmentations, such as adding noise, amplitude scaling, and temporal jitter, were applied to enhance the diversity of training data. The detailed implementation of these augmentations is provided in Appendix~\ref{appendix: A}.

\subsection{Spike Identification}
\label{subsec: method-Spike Identification}
We formulate spike identification as embedding each identified spike waveform $X_i \in \mathbb{R}^{L \times C}$, the voltage trace (waveform) of spike $i$, into a latent space to learn robust representations using contrastive learning. Then the learned embeddings are clustered into several groups, indicating putative neurons.

\begin{wrapfigure}{r}{0.6\textwidth}
 \vspace{-5mm}
  \begin{center}
    \includegraphics[width=1.0\linewidth]{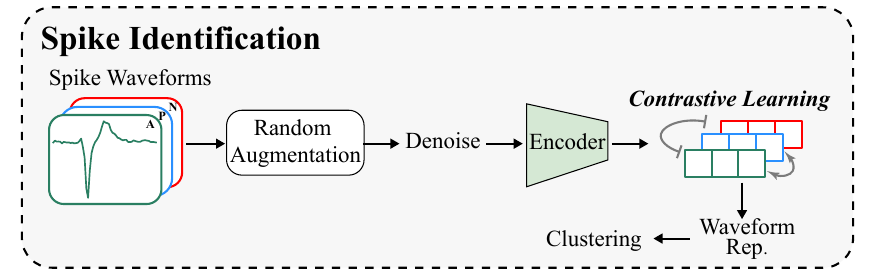}
  \end{center}
    \caption{
    Spike identification model in SimSort.
    }
    \label{fig:identification_model}
\end{wrapfigure}

The spike identification model was trained on the simulated spike waveform dataset (Sec.~\ref{subsec: Dataset Preparation}). The model consisted of several key components, as illustrated in Fig.~\ref{fig:identification_model} and described as follows.

\textbf{Contrastive Learning:}
We leveraged contrastive learning to learn representations that capture the relationships between spike waveforms (Fig.~\ref{fig:identification_model}). Contrastive learning sought to map neural data into an embedding space where related examples were positioned nearby, while unrelated examples were placed further apart \cite{dorkenwaldMultilayeredMapsNeuropil2023, yuVivoCelltypeBrain2024}. For each waveform $X_i$, we constructed a triplet by selecting $X_i^+$ with the same label and $X_i^-$ with a different label. The objective was \cite{schroffFaceNetUnifiedEmbedding2015}:
\begin{equation}
\mathcal{L}_{\text{triplet}} = \max \big( 0, 
\| f(X_i) - f(X_i^+)\|_2^2 
- \| f(X_i) - f(X_i^-)\|_2^2 + \alpha \big),
\end{equation}
where $f(X_i)$ denoted the embedding of $X_i$, and $\alpha$ was a margin enforcing separation between positive-negative pairs. Additionally, we uniformly applied data augmentation to each triplet waveforms. The augmentation methods followed those in Sec.~\ref{subsec: Data Augmentation}, with adjusted parameters (Appendix~\ref{appendix: A}) for better waveform representation, ensuring robustness to noise, amplitude variations, and temporal jitter.

\textbf{Denoising:}
The augmented waveform $X_i' \in \mathbb{R}^{L \times C}$ then underwent a denoising step using Singular Value Decomposition (SVD) \cite{eckartApproximationOneMatrix1936, cunninghamDimensionalityReductionLargescale2014}. The waveform was first reshaped and decomposed as $X_i' = U \Sigma V^\top$, where $U$, $\Sigma$, and $V$ represented the singular vectors and singular values. The first $k$ components were retained to reconstruct the waveform.

\textbf{Encoder:}
The denoised waveform $X_{i,\text{recon}}'$ was fed into a Gated Recurrent Unit (GRU) encoder \cite{choLearningPhraseRepresentations2014} (Fig.~\ref{fig:identification_model}), which processed the input sequentially. The final hidden state $h_T$ served as the learned representation $f(X_i)$, encapsulating the spatiotemporal features of the waveform for subsequent clustering tasks.

\textbf{Clustering:}
After obtaining the waveform representation, we applied dimensionality reduction using UMAP \cite{mcinnesUMAPUniformManifold2020} before clustering (Fig.~\ref{fig:identification_model}). We tested two clustering approaches: a parametric algorithm, Gaussian Mixture Model (GMM) \cite{dempsterMaximumLikelihoodIncomplete1977}, and a non-parametric one, Mean Shift (MS) \cite{chengMeanShiftMode1995}.

\subsection{Overall Spike Sorting Pipeline}
We established a cascaded spike sorting pipeline (Fig.~\ref{fig:pipeline}). During inference, raw multi-channel signals were processed through spike detection, waveform extraction, embedding generation, and clustering, yielding distinct neuronal units and their spike timestamps.

\section{Results}

With the large-scale dataset and a pretraining framework, we try to answer the following research questions (RQ).\\
\textbf{RQ1:} Can the model trained on our simulated dataset generalize to real-world spike sorting tasks? \\
\textbf{RQ2:} If the model performs well, what are the crucial underlying components? \\
\textbf{RQ3:} Does data scale hold the key of effective spike sorting?


\subsection{Benchmarks}
\label{subsec: Experimental Setup}

Since the models of SimSort has been trained on our simulated dataset, readers must be curious about how well it can generalize to real-world spike sorting tasks. To empirically investigate this problem, we respectively evaluate SimSort in the following benchmarks:
\begin{itemize}
    \item \textbf{Simulated test dataset} (with ground truth): BBP L6 dataset from our simulation (using different neuronal models from the training dataset).
    \item \textbf{Real-world data-based synthetic datasets} (with label): Hybrid \citep{maglandSpikeForestReproducibleWebfacing2020}, WaveClus \citep{martinezRealisticSimulationExtracellular2009}, IBL Neuropixels \citep{vishnubhotlaRobustGeneralizableRepresentations2023}. We evaluate both zero-shot (using only our simulated dataset) and fine-tuned (using also the dataset) performance of SimSort.
    \item \textbf{Real-world neural recording} (w/o ground truth, indirect validation): extracellular recordings collected from the mouse primary visual cortex during drifting grating stimulation.
\end{itemize}

The detailed information of benchmark datasets, hyper-parameters of models, and formulation of the evaluation metrics can be found in Appendix~\ref{appendix: A}.

\textbf{Evaluation Metrics}. To evaluate the performance of the overall spike sorting and spike detection tasks, we used Accuracy, Recall, and Precision as the evaluation metrics. Note that these metrics are specifically defined for the spike sorting context, where they measure the alignment between detected spikes and ground truth spike events, rather than their conventional definitions in classification tasks.
For the spike identification task, we used the Adjusted Rand Index (ARI) \cite{hubert1985comparing} to measure clustering accuracy relative to the ground truth. 

\begin{wraptable}{r}{7.5cm}
    \vspace{-6mm}
    \caption{
    \label{tab:bbp_L6_sorting}
        Spike sorting results on \textbf{BBP L6 dataset}.
        The results for other methods were obtained through SpikeInterface \cite{buccinoSpikeInterfaceUnifiedFramework2020}.
        Values are as mean $\pm$ S.E.M.  
    }
    \resizebox{0.5\columnwidth}{!}
    {
    \begin{tabular}{l|ccc}
    \toprule
    \multirow{2}{*}{\textbf{Methods}} & \multicolumn{3}{c}{\textbf{BBP L6
    \small(20 trials)}} \\
    & \textbf{Accuracy} & \textbf{Recall} & \textbf{Precision} \\
    
    \midrule
    KiloSort       & 0.49 $\pm$ 0.05 & 0.51 $\pm$ 0.05 & 0.53 $\pm$ 0.05 \\
    KiloSort2      & 0.51 $\pm$ 0.06 & 0.53 $\pm$ 0.06 & 0.53 $\pm$ 0.06 \\
    MountainSort4  & 0.84 $\pm$ 0.03 & 0.84 $\pm$ 0.03 & \textbf{0.93} $\pm$ \textbf{0.03} \\
    MountainSort5  & 0.66 $\pm$ 0.06 & 0.68 $\pm$ 0.06 & 0.79 $\pm$ 0.07 \\
    \rowcolor{simsortcolor!20} SimSort & \textbf{0.85} $\pm$ \textbf{0.02} & \textbf{0.90} $\pm$ \textbf{0.01} & \textbf{0.93} $\pm$ \textbf{0.02} \\
    \bottomrule
    \end{tabular}
    }
    \vspace{-3mm}
\end{wraptable}

\subsection{Spike Sorting Results on simulated dataset}
\label{subsec: Spike sorting results}
We first examined Simsort on a simulated test set (BBP L6). In Table~\ref{tab:bbp_L6_sorting}, we evaluated SimSort alongside widely used algorithms, including KiloSort \cite{pachitariuFastAccurateSpike2016}, KiloSort2, MountainSort4 \cite{chungFullyAutomatedApproach2017}, and MountainSort5, implemented via the SpikeInterface pipeline \cite{buccinoSpikeInterfaceUnifiedFramework2020}, on the simulated BBP L6 dataset. SimSort achieved the best performance across all metrics, demonstrating its robustness and strong zero-shot generalization ability under idealized, simulated conditions.

\begin{table*}[h]
    
    \centering
    \caption{
    \label{tab:hybrid_sorting_table}
        Spike sorting results on \textbf{Hybrid dataset}.
        Data for other methods were sourced from SpikeForest \cite{maglandSpikeForestReproducibleWebfacing2020}. 
        Values are mean $\pm$ S.E.M. Best results are in \textbf{bold}. 
        Note that SimSort did not use this dataset for training, indicating a zero-shot generalization.
    }
    \footnotesize
    \resizebox{0.9\linewidth}{!}{
    \begin{tabular}{l|ccc|ccc}
    \toprule
    \multirow{2}{*}{\textbf{Methods}} & \multicolumn{3}{c|}{\textbf{Hybrid-static \small(SNR$>$3, 9 recordings)}} 
                                      & \multicolumn{3}{c}{\textbf{Hybrid-drift \small(SNR$>$3, 9 recordings)}} \\
    & \textbf{Accuracy} & \textbf{Recall} & \textbf{Precision}
    & \textbf{Accuracy} & \textbf{Recall} & \textbf{Precision} \\
    \midrule
    
    HerdingSpikes2  & 0.35 $\pm$ 0.01 & 0.44 $\pm$ 0.02 & 0.53 $\pm$ 0.01
                    & 0.29 $\pm$ 0.01 & 0.37 $\pm$ 0.02 & 0.48 $\pm$ 0.02 \\
    IronClust       & 0.57 $\pm$ 0.04 & \textbf{0.81 $\pm$ 0.01} & 0.60 $\pm$ 0.04
                    & 0.54 $\pm$ 0.03 & 0.71 $\pm$ 0.02 & 0.65 $\pm$ 0.03 \\
    JRClust         & 0.47 $\pm$ 0.04 & 0.63 $\pm$ 0.02 & 0.59 $\pm$ 0.03
                    & 0.35 $\pm$ 0.03 & 0.48 $\pm$ 0.03 & 0.57 $\pm$ 0.02 \\
    KiloSort        & 0.60 $\pm$ 0.02 & 0.65 $\pm$ 0.02 & 0.72 $\pm$ 0.02
                    & 0.51 $\pm$ 0.02 & 0.62 $\pm$ 0.01 & \textbf{0.72} $\pm$ \textbf{0.03} \\
    KiloSort2       & 0.39 $\pm$ 0.03 & 0.37 $\pm$ 0.03 & 0.51 $\pm$ 0.03
                    & 0.30 $\pm$ 0.02 & 0.31 $\pm$ 0.02 & 0.57 $\pm$ 0.04 \\
    MountainSort4   & 0.59 $\pm$ 0.02 & 0.73 $\pm$ 0.01 & 0.73 $\pm$ 0.03
                    & 0.36 $\pm$ 0.02 & 0.57 $\pm$ 0.02 & 0.61 $\pm$ 0.03 \\
    SpykingCircus   & 0.57 $\pm$ 0.01 & 0.63 $\pm$ 0.01 & 0.75 $\pm$ 0.03
                    & 0.48 $\pm$ 0.02 & 0.55 $\pm$ 0.02 & 0.68 $\pm$ 0.03 \\
    Tridesclous     & 0.54 $\pm$ 0.03 & 0.66 $\pm$ 0.02 & 0.59 $\pm$ 0.04
                    & 0.37 $\pm$ 0.02 & 0.52 $\pm$ 0.03 & 0.55 $\pm$ 0.04 \\
    \rowcolor{simsortcolor!20} SimSort& \textbf{0.62} $\pm$ \textbf{0.04} & 0.68 $\pm$ 0.04 & \textbf{0.77} $\pm$ \textbf{0.03}
                                       & \textbf{0.56} $\pm$ \textbf{0.03} & \textbf{0.63} $\pm$ \textbf{0.03} & 0.69 $\pm$ 0.03\\
    \bottomrule
    \end{tabular}
    }
    \vspace{-4mm}
\end{table*}

\begin{table*}[h]
    \centering
    \caption{
    \label{tab:waveclus_sorting_table}
        Spike sorting results on \textbf{WaveClus dataset}.
        The results for other methods were obtained through SpikeInterface \cite{buccinoSpikeInterfaceUnifiedFramework2020}.
        Values are mean $\pm$ S.E.M.  
        Note that SimSort did not use this dataset for training, indicating a zero-shot generalization. 
    }
    \vspace{1mm}
    \footnotesize
    \resizebox{0.9\linewidth}{!}{
    \begin{tabular}{l|ccc|ccc}
    \toprule
    \multirow{2}{*}{\textbf{Methods}} & \multicolumn{3}{c|}{\textbf{Easy \small(12 recordings)}} 
                                      & \multicolumn{3}{c}{\textbf{Difficult \small (8 recordings)}} \\
    & \textbf{Accuracy} & \textbf{Recall} & \textbf{Precision}
    & \textbf{Accuracy} & \textbf{Recall} & \textbf{Precision} \\
    \midrule
    
    KiloSort        & 0.54 $\pm$ 0.12 & 0.46 $\pm$ 0.15 & 0.62 $\pm$ 0.13
                    & 0.17 $\pm$ 0.07 & 0.23 $\pm$ 0.10 & 0.18 $\pm$ 0.07 \\
    KiloSort2       & 0.61 $\pm$ 0.09 & 0.66 $\pm$ 0.09 & 0.67 $\pm$ 0.09
                    & 0.10 $\pm$ 0.07 & 0.12 $\pm$ 0.09 & 0.10 $\pm$ 0.07 \\
    MountainSort4   & 0.73 $\pm$ 0.08 & 0.79 $\pm$ 0.07 & 0.79 $\pm$ 0.08
                    & 0.48 $\pm$ 0.16 & 0.49 $\pm$ 0.16 & 0.52 $\pm$ 0.17 \\
    MountainSort5   & 0.71 $\pm$ 0.09 & 0.74 $\pm$ 0.08 & 0.79 $\pm$ 0.10
                    & 0.42 $\pm$ 0.15 & 0.46 $\pm$ 0.15 & 0.46 $\pm$ 0.16 \\
                    
    \rowcolor{simsortcolor!20} SimSort & \textbf{0.75} $\pm$ \textbf{0.06} & \textbf{0.84} $\pm$ \textbf{0.04} & \textbf{0.85} $\pm$ \textbf{0.05}
                                       & \textbf{0.71} $\pm$ \textbf{0.10} & \textbf{0.81} $\pm$ \textbf{0.07} & \textbf{0.81} $\pm$ \textbf{0.09} \\
    \bottomrule
    \end{tabular}
    }
    \vspace{-4mm}
\end{table*}

\subsection{Spiking Sorting results on real-world data-based synthetic datasets}
\paragraph{zero-shot generalization}
Given SimSort’s strong performance on simulated data, we extended our evaluation to real-world data-based datasets to further test its zero-shot generalizability and robustness (RQ1). On the Hybrid dataset, we compared SimSort against several algorithms via SpikeForest, including HerdingSpikes2 \cite{hilgenUnsupervisedSpikeSorting2017}, IronClust, JRClust, KiloSort \cite{pachitariuFastAccurateSpike2016}, KiloSort2, MountainSort4 \cite{chungFullyAutomatedApproach2017}, SpykingCircus, and Tridesclous. The results are summarized in Table~\ref{tab:hybrid_sorting_table} and Fig.~\ref{fig: hybrid_waveclus-acc}. On this dataset, SimSort demonstrated the highest Accuracy and Precision in static recordings and achieved the best Accuracy and Recall under drift conditions, although it slightly trailed KiloSort in Precision.

For WaveClus dataset, we used SpikeInterface pipeline to run KiloSort, KiloSort2, MountainSort4, and MountainSort5. Results are provided in Table~\ref{tab:waveclus_sorting_table} and Fig.~\ref{fig: hybrid_waveclus-acc}. SimSort performed reliably across both easy and difficult subsets, showing a pronounced advantage on the challenging ``difficult'' set. This performance was likely due to its ability to handle the unique challenges of this subset, such as lower SNRs and reduced inter-class waveform variability, which resulted in significant overlaps between units. These results underscored SimSort's effectiveness in noisy and complex scenarios.

\paragraph{Fine-tuning}
Beyond demonstrating SimSort's zero-shot transfer ability on unseen datasets, we also investigate its performance after fine-tuning on a small amount of data from the target domain (Table~\ref{tab: finetuned_experiments}). While the overall improvements in sorting metrics are modest, we observe notable changes in spike detection performance, particularly in the balance between Precision and Recall. These changes suggest that fine-tuning helps the model adapt to dataset-specific waveform and noise characteristics, enhancing its practicality with minimal labeled data.

\begin{table}[h]
    \vspace{-4mm}
    \centering
    \caption{
    \label{tab: finetuned_experiments}
    Performance metrics of SimSort after fine-tuning on different datasets.The hybrid-static and hybrid-drift datasets were fine-tuned on recordings 1-3 and tested on recordings 4-9 (train/test set split). The waveclus -easy and -difficult datasets were fine-tuned on subsets easy1/diff1 and tested on easy2/diff2, respectively. The IBL Neuropixels dataset was fine-tuned on 100 neurons and evaluated on 50 test sets, each with 10 randomly selected neurons.
    }
    \vspace{2mm}
    \resizebox{1.0\columnwidth}{!}
    {
    \begin{tabular}{llccc}
        \toprule
        Dataset & Model & Spike Detection (Acc/Rec/Prec) & Spike Identification (ARI) & Spike Sorting (Acc/Rec/Prec) \\
        \midrule
        \multirow{2}{*}{Hybrid-static} & SimSort & 0.71±0.02 / 0.83±0.02 / 0.83±0.02 & 0.90±0.02 & 0.59±0.03 / 0.65±0.03 / 0.76±0.02 \\
        & SimSort-FT & 0.75±0.03 / 0.89±0.02 / 0.82±0.02 & 0.92±0.02 & 0.58±0.03 / 0.65±0.03 / 0.73±0.02 \\
        \midrule
        \multirow{2}{*}{Hybrid-drift} & SimSort & 0.67±0.02 / 0.80±0.02 / 0.81±0.01 & 0.88±0.03 & 0.54±0.09 / 0.61±0.09 / 0.67±0.07 \\
        & SimSort-FT & 0.73±0.03 / 0.85±0.03 / 0.83±0.01 & 0.88±0.02 & 0.56±0.04 / 0.63±0.03 / 0.69±0.03 \\
        \midrule
        \multirow{2}{*}{Waveclus-easy} & SimSort & 0.59±0.02 / 0.99±0.00 / 0.59±0.02 & 0.91±0.01 & 0.70±0.11 / 0.79±0.07 / 0.84±0.09 \\
        & SimSor-FT & 0.94±0.01 / 0.96±0.00 / 0.98±0.01 & 0.91±0.01 & 0.73±0.08 / 0.79±0.05 / 0.91±0.04 \\
        \midrule
        \multirow{2}{*}{Waveclus-difficult} & SimSort & 0.66±0.08 / 0.99±0.00 / 0.67±0.08 & 0.88±0.01 & 0.76±0.14 / 0.84±0.09 / 0.84±0.12 \\
        & SimSort-FT & 0.95±0.01 / 0.96±0.00 / 0.99±0.01 & 0.87±0.03 & 0.77±0.14 / 0.82±0.10 / 0.86±0.13 \\
        \midrule
        \multirow{2}{*}{IBL Neuropixels} & SimSort & N/A & 0.49±0.09 & N/A \\
        & SimSort-FT & N/A & 0.53±0.09 & N/A \\
        \bottomrule
    \end{tabular}
    }
\end{table}

\subsection{Real-World Neural Recording}
\label{subsec: real_world_application}

\begin{figure*}[h]
    \centering
    \includegraphics[width=\linewidth]{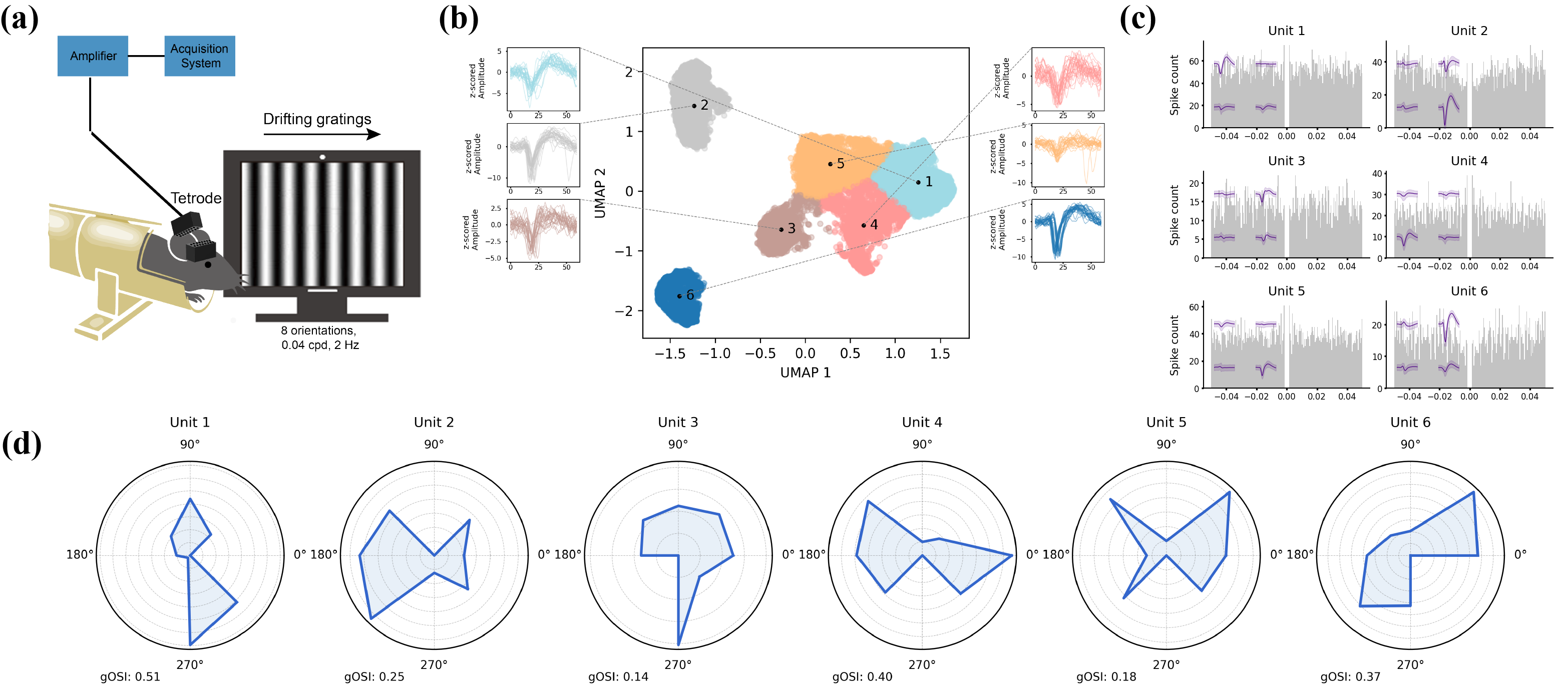}
    \vspace{-2mm}
    \caption{
    Evaluation of SimSort on real experimental data (See Fig.~\ref{fig: realdata_sorting_visualization} for additional recording results).
    (a) Experimental setup: extracellular recordings from the mouse primary visual cortex during drifting gratings in eight directions.
    (b) Visualization of SimSort's sorting results.
    (c) Autocorrelograms and average waveform of each unit.
    (d) Orientation tuning curves and global orientation selectivity index (gOSI) for each unit.
    \label{fig: real_world_experimental_task}
    } 
\end{figure*}

Beyond quantitative benchmarks, we further assessed SimSort in a real experimental setting. Specifically, we applied it to extracellular recordings (collected in our wet lab, see Appendix~\ref{appendix: test datasets}) from the mouse primary visual cortex obtained using tetrodes during presentation of full-field drifting gratings in 8 orientations (Fig.~\ref{fig: real_world_experimental_task}a). From a representative recording session, SimSort identified six putative single units (Fig.~\ref{fig: real_world_experimental_task}b). These units exhibited distinct spike waveforms and their autocorrelograms showed clear refractory period structure(Fig.~\ref{fig: real_world_experimental_task}c), suggesting good isolation. We further analyzed the orientation tuning of these units. Several exhibited selective responses, with global orientation selectivity index (gOSI) values exceeding 0.2 (Fig.~\ref{fig: real_world_experimental_task}d; Appendix~\ref{appendix: A} for definition). Across 10 tetrode recordings, SimSort isolated a total of 40 units, with a mean gOSI of 0.27 ± 0.14 (mean ± S.D., Fig.~\ref{fig: realdata_sorting_visualization}). These results demonstrate the practical applicability of SimSort to real neural recordings.

\subsection{Detailed Analysis}
To explain why our framework achieved higher overall spike sorting accuracy (RQ2), we dissected SimSort and analyzed each component, i.e., spike detection and spike identification models. Furthermore, we presented some case studies of the latent embedding of spikes from multiple neurons.

\label{subsec: Detailed_experiment_analysis}
\vspace{-3mm}
\paragraph{Spike Detection}
\label{sec: comparasion of detection}
We first compared our detection model to a commonly used threshold-based detection method (see Appendix~\ref{appendix: A} for details). The results are shown in Table~\ref{tab:hybrid_detection_table}. SimSort's spike detector consistently outperformed the threshold-based method across all metrics on the Hybrid dataset. In the static condition, SimSort improved accuracy by $\sim$18$\%$, recall by $\sim$18$\%$, and precision by $\sim$1$\%$. Under the drift condition, the improvements were $\sim$13$\%$ in accuracy, $\sim$17$\%$ in recall, and $\sim$1$\%$ in precision. It is important to note that the performance of the Threshold detector was highly sensitive to the choice of threshold, with results varying significantly across different settings. The results shown here were obtained under the best-performing thresholds. In contrast, SimSort achieved its superior performance without manual tuning, demonstrating its robustness and adaptability to varying recording conditions. These results indicated the effectiveness of our detection model, providing a solid foundation for the overall performance of SimSort in spike sorting tasks.

\begin{table*}[ht]
  \caption{Spike detection and identification results on the \textbf{Hybrid dataset}. Values are mean $\pm$ S.E.M. }
  \centering
  \begin{subtable}[t]{0.48\textwidth}
    \centering
    \caption{Spike detection results. The threshold detector used the best-performing threshold by grid search.}
    \label{tab:hybrid_detection_table}
    \resizebox{\linewidth}{!}{%
      \begin{tabular}{lcccc} 
        \toprule
        \textbf{Dataset} & \textbf{Method} & \textbf{Accuracy} & \textbf{Recall} & \textbf{Precision} \\
        \midrule
            & Threshold detector           & 0.61$\pm$0.03 & 0.71$\pm$0.02 & 0.81$\pm$0.02 \\
        \rowcolor{simsortcolor!20}
        \multirow{2}{*}[12pt]{Static} & SimSort's detector & \textbf{0.72$\pm$0.03} & \textbf{0.84$\pm$0.02} & \textbf{0.82$\pm$0.02} \\
        \midrule
            & Threshold detector           & 0.60$\pm$0.03 & 0.70$\pm$0.03 & 0.80$\pm$0.02 \\
        \rowcolor{simsortcolor!20}
        \multirow{2}{*}[12pt]{Drift} & SimSort's detector & \textbf{0.68$\pm$0.03} & \textbf{0.82$\pm$0.02} & \textbf{0.81$\pm$0.02} \\
        \bottomrule
      \end{tabular}%
    }
  \end{subtable}%
  \hfill
  \begin{subtable}[t]{0.48\textwidth}
  \centering
  \caption{Spike identification results. All results averaged across 20 repeats of GMM.}
  \label{tab:hybrid_identification_table}
  \resizebox{\linewidth}{!}{%
    \begin{tabular}{lcccc}
      \toprule
      \textbf{Dataset} & \textbf{PCA} & \textbf{$t$-SNE} & \textbf{UMAP} & \textbf{Ours} \\
      \midrule
      Hybrid-static & 0.79 $\pm$ 0.04 & 0.88 $\pm$ 0.02 & 0.88 $\pm$ 0.02 & \textbf{0.91} $\pm$ \textbf{0.02} \\
      Hybrid-drift  & 0.74 $\pm$ 0.04 & 0.85 $\pm$ 0.02 & 0.85 $\pm$ 0.01 & \textbf{0.89} $\pm$ \textbf{0.03} \\
      \bottomrule                                   
    \end{tabular}%
  }
\end{subtable}
\end{table*}

    
    

\paragraph{Spike Identification}
After detecting spikes, our spike identification model extracted waveform representations in the latent space, enabling subsequent clustering. To effectively demonstrate the superiority of our identification model, we assembled a waveform dataset by gathering waveform data from each neuron in every hybrid recording. Using this dataset, we compared the 2D UMAP embedding of our model's inferred representations with features extracted directly from waveforms using widely adopted dimensionality reduction techniques, including \textbf{PCA} \cite{FRS1901LIIIOL}, $\boldsymbol{t}$\textbf{-SNE} \cite{maatenVisualizingDataUsing2008}, and \textbf{UMAP} \cite{mcinnesUMAPUniformManifold2020}. For evaluation, we applied a parametric clustering method, \textbf{GMM} \cite{dempsterMaximumLikelihoodIncomplete1977}, to compute the Adjusted Rand Index (ARI), providing a quantitative measure of the reliability of the representations.

    

As shown in Table~\ref{tab:hybrid_identification_table}, SimSort substantially outperformed these conventional methods under both static and drift conditions, achieving the highest average ARI values of \textbf{0.91} and \textbf{0.89}, respectively. These results revealed the robustness and effectiveness of our approach in deriving meaningful and reliable representations from extracellular data.

Additionally, we compared SimSort with CEED \cite{vishnubhotlaRobustGeneralizableRepresentations2023}, a recent framework for representation learning on extracellular data, using the author-provided model. Note that CEED is trained on datasets derived from IBL DanLab DY016 and DY009 Neuropixels recordings, which are more closely related to our test sets. We evaluated both SimSort and CEED on fifty test sets from the \textbf{IBL Neuropixels dataset} (specifically derived from the IBL CortexLab KS046 recording). As presented in Table~\ref{tab:IBL_identification_table}, Fig.~\ref{fig: IBL_bar},~\ref{fig: IBL_comparisions}, our method, without training on the dataset, achieved performance comparable to CEED.

\begin{figure*}[th]
    \centering
    \includegraphics[width=1\linewidth]{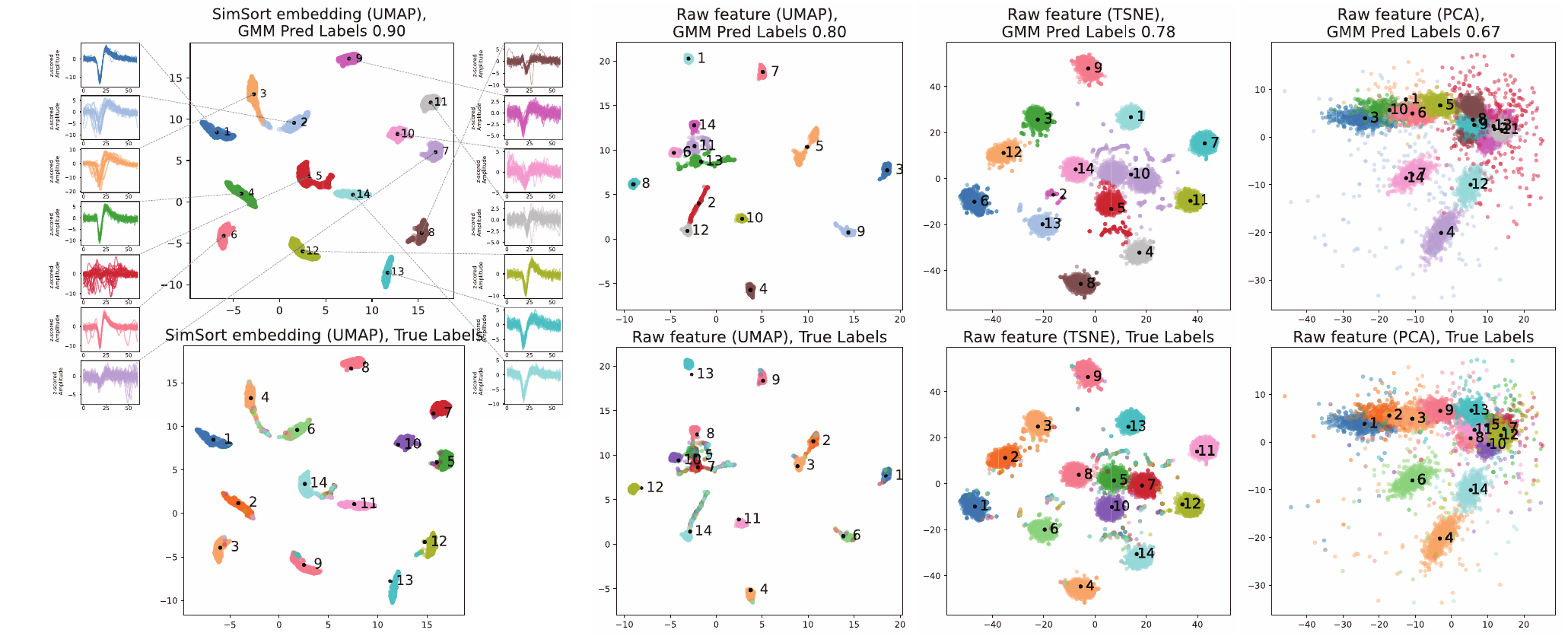}
    \vspace{-2mm}
    \caption{
    \label{fig: hybrid_identification_example}
        Visualized results of spike identification on an example recording in the Hybrid static subset. 
        See Appendix \ref{appendix: more showcases} for visualization results for additional samples.
    }
    \vspace{-3mm}
\end{figure*}

\textbf{Case Studies:}
In Fig.~\ref{fig: hybrid_identification_example}, we visualized the results of spike identification on an example recording in the Hybrid static subset. SimSort's embeddings (first column) demonstrated superior clustering performance, with a GMM clustering ARI score of 0.90, closely aligning with true labels. In contrast, UMAP, $t$-SNE, and PCA embeddings (second to fourth columns) yielded lower clustering scores of 0.80, 0.78, and 0.67, respectively. SimSort embeddings showed clear separation between neuron clusters, while other methods exhibited significant overlap, highlighting the advantages of SimSort in learning robust and discriminative features.

\textbf{Impact of Dimensionality Reduction and Clustering Methods:}
To evaluate the adaptability of SimSort, we conducted a study to assess the impact of different dimensionality reduction techniques (PCA, $t$-SNE, UMAP, and None) and clustering algorithms (GMM, KMeans \cite{macqueenMethodsClassificationAnalysis1967}, Mean Shift \cite{chengMeanShiftMode1995}, and Spectral Clustering \cite{ngSpectralClusteringAnalysis2001}). Dimensionality reduction was applied to latent features extracted from waveforms, with an additional setup where clustering was performed directly on the raw latent features without dimensionality reduction (None). As illustrated in Fig.~\ref{fig: ablation_study}, the combination of UMAP and Mean Shift achieved the highest accuracy on both hybrid static and drift subsets. Across many other combinations, SimSort also maintained competitive performance, demonstrating its flexibility and robustness to various configurations.

\subsection{Scaling Law of Data Size}
\label{subsec: scaling law}
\begin{wrapfigure}{r}{6.8cm}
    \vspace{-9mm}
    \includegraphics[width=6.2cm]{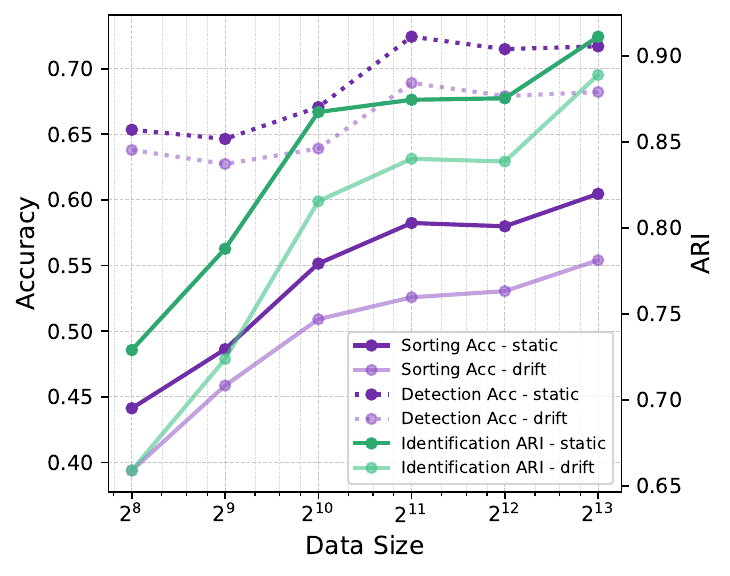}
    \caption{
    \label{fig: scaling_law}
        Performance scaling with increasing training data size on the Hybrid-drift and -static.
    }
    \vspace{-5mm}
\end{wrapfigure}

As we emphasized the importance of a large-scale dataset throughout this work, it was interesting to evaluate how SimSort's performance scaled with the size of the training dataset (RQ3). Specifically, we evaluated performance across sorting accuracy, detection accuracy, and identification ARI score, capturing how each metric varied with the growing size of the training dataset, which ranged from $2^8$ to $2^{13}$ trials. As Fig.~\ref{fig: scaling_law} shows, larger datasets led to almost consistent performance improvements on both hybrid static and drift subsets, reflecting that the large-scale dataset holds the key of SimSort's effectiveness.

\section{Conclusion}
In this paper, we presented SimSort, a data-driven framework for automated spike sorting, built upon a biologically realistic large-scale extracellular dataset. SimSort achieves competitive or superior performance compared to commonly used spike sorting algorithms on both simulated and real-world benchmarks, demonstrating its robustness and scalability. Evaluations further highlighted its zero-shot transfer capability.

SimSort has limitations. The models were trained on 4-channel tetrode data. While SimSort shows promising performance on tasks with high-density probes (IBL Neuropixels), future work should extend the training with diverse electrode geometries.

\begin{ack}
This work was supported by AI for Science Foundation of Fudan University (FudanX24AI046 to Y.G.) and Microsoft Research.
\end{ack}


\bibliographystyle{unsrt}
\bibliography{arxiv}

@article{lewickiReviewMethodsSpike1998,
  title = {A Review of Methods for Spike Sorting: The Detection and Classification of Neural Action Potentials},
  author = {Lewicki, Michael S.},
  year = {1998},
  journal = {Network: Computation in Neural Systems},
  volume = {9},
  number = {4},
  pages = {R53}
}

@article{buzsakiLargescaleRecordingNeuronal2004,
  title = {Large-Scale Recording of Neuronal Ensembles},
  author = {Buzsáki, György},
  year = {2004},
  journal = {Nature Neuroscience},
  volume = {7},
  number = {5},
  pages = {446--451},
  publisher = {Nature Publishing Group}
}

@article{quianquirogaExtractingInformationNeuronal2009,
  title = {Extracting Information from Neuronal Populations: Information Theory and Decoding Approaches},
  author = {Quian Quiroga, Rodrigo and Panzeri, Stefano},
  year = {2009},
  journal = {Nature Reviews Neuroscience},
  volume = {10},
  number = {3},
  pages = {173--185},
  publisher = {Nature Publishing Group}
}

@article{martinezRealisticSimulationExtracellular2009,
  title = {Realistic Simulation of Extracellular Recordings},
  author = {Martinez, Juan and Pedreira, Carlos and Ison, Matias J. and Quian Quiroga, Rodrigo},
  date = {2009-11-15},
  journaltitle = {Journal of Neuroscience Methods},
  shortjournal = {Journal of Neuroscience Methods},
  volume = {184},
  number = {2},
  pages = {285--293},
  url = {https://www.sciencedirect.com/science/article/pii/S0165027009004506}
}

@inproceedings{sarkarAdvancedSpikeSorting2024,
  title = {Advanced {{Spike Sorting Approaches}} in {{Implantable VLSI Wireless Brain Computer Interfaces}}: {{A Survey}}},
  booktitle = {2024 {{IEEE Region}} 10 {{Symposium}} ({{TENSYMP}})},
  author = {Sarkar, Soujatya},
  year = {2024},
  pages = {1--7}
}

@article{joshiDynamicSynchronizationHippocampal2023,
  title = {Dynamic Synchronization between Hippocampal Representations and Stepping},
  author = {Joshi, Abhilasha and Denovellis, Eric L. and Mankili, Abhijith and Meneksedag, Yagiz and Davidson, Thomas J. and Gillespie, Anna K. and Guidera, Jennifer A. and Roumis, Demetris and Frank, Loren M.},
  year = {2023},
  journal = {Nature},
  volume = {617},
  number = {7959},
  pages = {125--131},
  publisher = {Nature Publishing Group}
}

@article{chungFullyAutomatedApproach2017,
  title = {A {{Fully Automated Approach}} to {{Spike Sorting}}},
  author = {Chung, Jason E. and Magland, Jeremy F. and Barnett, Alex H. and Tolosa, Vanessa M. and Tooker, Angela C. and Lee, Kye Y. and Shah, Kedar G. and Felix, Sarah H. and Frank, Loren M. and Greengard, Leslie F.},
  year = {2017},
  journal = {Neuron},
  volume = {95},
  number = {6},
  pages = {1381-1394.e6}
}

@article{steinmetzNeuropixels20Miniaturized2021a,
  title = {Neuropixels 2.0: {{A}} Miniaturized High-Density Probe for Stable, Long-Term Brain Recordings},
  author = {Steinmetz, Nicholas A. and Aydin, Cagatay and Lebedeva, Anna and Okun, Michael and Pachitariu, Marius and Bauza, Marius and Beau, Maxime and Bhagat, Jai and Böhm, Claudia and Broux, Martijn and Chen, Susu and Colonell, Jennifer and Gardner, Richard J. and Karsh, Bill and Kloosterman, Fabian and Kostadinov, Dimitar and {Mora-Lopez}, Carolina and O’Callaghan, John and Park, Junchol and Putzeys, Jan and Sauerbrei, Britton and {van Daal}, Rik J. J. and Vollan, Abraham Z. and Wang, Shiwei and Welkenhuysen, Marleen and Ye, Zhiwen and Dudman, Joshua T. and Dutta, Barundeb and Hantman, Adam W. and Harris, Kenneth D. and Lee, Albert K. and Moser, Edvard I. and O’Keefe, John and Renart, Alfonso and Svoboda, Karel and Häusser, Michael and Haesler, Sebastian and Carandini, Matteo and Harris, Timothy D.},
  year = {2021},
  journal = {Science},
  volume = {372},
  number = {6539},
  pages = {eabf4588}
}

@article{chungHighDensityLongLastingMultiregion2019,
  title = {High-{{Density}}, {{Long-Lasting}}, and {{Multi-region Electrophysiological Recordings Using Polymer Electrode Arrays}}},
  author = {Chung, Jason E. and Joo, Hannah R. and Fan, Jiang Lan and Liu, Daniel F. and Barnett, Alex H. and Chen, Supin and {Geaghan-Breiner}, Charlotte and Karlsson, Mattias P. and Karlsson, Magnus and Lee, Kye Y. and Liang, Hexin and Magland, Jeremy F. and Pebbles, Jeanine A. and Tooker, Angela C. and Greengard, Leslie F. and Tolosa, Vanessa M. and Frank, Loren M.},
  year = {2019},
  journal = {Neuron},
  volume = {101},
  number = {1},
  pages = {21-31.e5},
  publisher = {Elsevier}
}

@article{hongNovelElectrodeTechnologies2019,
  title = {Novel Electrode Technologies for Neural Recordings},
  author = {Hong, Guosong and Lieber, Charles M.},
  year = {2019},
  journal = {Nature Reviews Neuroscience},
  volume = {20},
  number = {6},
  pages = {330--345},
  publisher = {Nature Publishing Group}
}

@article{lefebvreRecentProgressMultielectrode2016,
  title = {Recent Progress in Multi-Electrode Spike Sorting Methods},
  author = {Lefebvre, Baptiste and Yger, Pierre and Marre, Olivier},
  year = {2016},
  journal = {Journal of Physiology-Paris},
  series = {{{SI}}: {{GDR Multielectrode}}},
  volume = {110},
  number = {4, Part A},
  pages = {327--335}
}

@article{sibilleHighdensityElectrodeRecordings2022,
  title = {High-Density Electrode Recordings Reveal Strong and Specific Connections between Retinal Ganglion Cells and Midbrain Neurons},
  author = {Sibille, Jérémie and Gehr, Carolin and Benichov, Jonathan I. and Balasubramanian, Hymavathy and Teh, Kai Lun and Lupashina, Tatiana and Vallentin, Daniela and Kremkow, Jens},
  year = {2022},
  journal = {Nature Communications},
  volume = {13},
  number = {1},
  pages = {5218},
  publisher = {Nature Publishing Group}
}

@article{liuDecodingCognitionSpontaneous2022,
  title = {Decoding Cognition from Spontaneous Neural Activity},
  author = {Liu, Yunzhe and Nour, Matthew M. and Schuck, Nicolas W. and Behrens, Timothy E. J. and Dolan, Raymond J.},
  year = {2022},
  journal = {Nature Reviews Neuroscience},
  volume = {23},
  number = {4},
  pages = {204--214},
  publisher = {Nature Publishing Group}
}

@article{park128ChannelFPGABasedRealTime2017,
  title = {A 128-{{Channel FPGA-Based Real-Time Spike-Sorting Bidirectional Closed-Loop Neural Interface System}}},
  author = {Park, Jongkil and Kim, Gookhwa and Jung, Sang-Don},
  year = {2017},
  journal = {IEEE Transactions on Neural Systems and Rehabilitation Engineering},
  volume = {25},
  number = {12},
  pages = {2227--2238}
}

@inproceedings{hanLiveDemonstrationEfficient2024,
  title = {Live {{Demonstration}}: {{An Efficient Hardware}} for {{Real-time Multi-channel Spike Sorting}} with {{Localization}}},
  booktitle = {2024 {{IEEE Biomedical Circuits}} and {{Systems Conference}} ({{BioCAS}})},
  author = {Han, Yuntao and Wang, Shiwei and Hamilton, Alister},
  year = {2024},
  pages = {1--1}
}

@article{quirogaUnsupervisedSpikeDetection2004,
  title = {Unsupervised Spike Detection and Sorting with Wavelets and Superparamagnetic Clustering},
  author = {Quiroga, R.Q. and Nadasdy, Z. and {Ben-Shaul}, Y.},
  year = {2004},
  journal = {Neural Computation},
  volume = {16},
  number = {8},
  pages = {1661--1687}
}

@article{rossantSpikeSortingLarge2016,
  title = {Spike Sorting for Large, Dense Electrode Arrays},
  author = {Rossant, Cyrille and Kadir, Shabnam N. and Goodman, Dan F. M. and Schulman, John and Hunter, Maximilian L. D. and Saleem, Aman B. and Grosmark, Andres and Belluscio, Mariano and Denfield, George H. and Ecker, Alexander S. and Tolias, Andreas S. and Solomon, Samuel and Buzsáki, György and Carandini, Matteo and Harris, Kenneth D.},
  year = {2016},
  journal = {Nature Neuroscience},
  volume = {19},
  number = {4},
  pages = {634--641},
  publisher = {Nature Publishing Group}
}

@inproceedings{pachitariuFastAccurateSpike2016,
  title = {Fast and Accurate Spike Sorting of High-Channel Count Probes with {{KiloSort}}},
  booktitle = {Advances in {{Neural Information Processing Systems}}},
  author = {Pachitariu, Marius and Steinmetz, Nicholas A and Kadir, Shabnam N and Carandini, Matteo and Harris, Kenneth D},
  year = {2016},
  volume = {29},
  publisher = {Curran Associates, Inc.}
}

@article{pachitariuSpikeSortingKilosort42024,
  title = {Spike Sorting with {{Kilosort4}}},
  author = {Pachitariu, Marius and Sridhar, Shashwat and Pennington, Jacob and Stringer, Carsen},
  year = {2024},
  journal = {Nature Methods},
  volume = {21},
  number = {5},
  pages = {914--921},
  publisher = {Nature Publishing Group}
}

@article{hilgenUnsupervisedSpikeSorting2017,
  title = {Unsupervised {{Spike Sorting}} for {{Large-Scale}}, {{High-Density Multielectrode Arrays}}},
  author = {Hilgen, Gerrit and Sorbaro, Martino and Pirmoradian, Sahar and Muthmann, Jens-Oliver and Kepiro, Ibolya Edit and Ullo, Simona and Ramirez, Cesar Juarez and Encinas, Albert Puente and Maccione, Alessandro and Berdondini, Luca and Murino, Vittorio and Sona, Diego and Zanacchi, Francesca Cella and Sernagor, Evelyne and Hennig, Matthias Helge},
  year = {2017},
  journal = {Cell Reports},
  volume = {18},
  number = {10},
  pages = {2521--2532},
  publisher = {Elsevier}
}

@inproceedings{leeYASSAnotherSpike2017,
  title = {{{YASS}}: {{Yet Another Spike Sorter}}},
  booktitle = {Advances in {{Neural Information Processing Systems}}},
  author = {Lee, Jin Hyung and Carlson, David E and Shokri Razaghi, Hooshmand and Yao, Weichi and Goetz, Georges A and Hagen, Espen and Batty, Eleanor and Chichilnisky, E.J. and Einevoll, Gaute T. and Paninski, Liam},
  year = {2017},
  volume = {30},
  publisher = {Curran Associates, Inc.}
}

@inproceedings{vishnubhotlaRobustGeneralizableRepresentations2023,
  title = {Towards Robust and Generalizable Representations of Extracellular Data Using Contrastive Learning},
  booktitle = {Thirty-Seventh {{Conference}} on {{Neural Information Processing Systems}}},
  author = {Vishnubhotla, Ankit and Loh, Charlotte and Srivastava, Akash and Paninski, Liam and Hurwitz, Cole Lincoln},
  year = {2023}
}

@article{buccinoSpikeSortingNew2022,
  title = {Spike Sorting: New Trends and Challenges of the Era of High-Density Probes},
  author = {Buccino, Alessio P. and Garcia, Samuel and Yger, Pierre},
  year = {2022},
  journal = {Progress in Biomedical Engineering},
  volume = {4},
  number = {2},
  pages = {022005},
  publisher = {IOP Publishing}
}

@article{FRS1901LIIIOL,
  title = {On Lines and Planes of Closest Fit to Systems of Points in Space},
  author = {Pearson, Karl},
  year = {1901},
  journal = {Philosophical Magazine Series},
  volume = {2},
  pages = {559--572}
}

@article{maatenVisualizingDataUsing2008,
  title = {Visualizing {{Data}} Using T-{{SNE}}},
  author = {van der Maaten, Laurens and Hinton, Geoffrey},
  year = {2008},
  journal = {Journal of Machine Learning Research},
  volume = {9},
  number = {86},
  pages = {2579--2605}
}

@article{veerabhadrappaCompatibilityEvaluationClustering2020,
  title = {Compatibility {{Evaluation}} of {{Clustering Algorithms}} for {{Contemporary Extracellular Neural Spike Sorting}}},
  author = {Veerabhadrappa, Rakesh and Ul Hassan, Masood and Zhang, James and Bhatti, Asim},
  year = {2020},
  journal = {Frontiers in Systems Neuroscience},
  volume = {14},
  publisher = {Frontiers}
}

@inproceedings{baldiAutoencodersUnsupervisedLearning2012,
  title = {Autoencoders, {{Unsupervised Learning}}, and {{Deep Architectures}}},
  booktitle = {Proceedings of {{ICML Workshop}} on {{Unsupervised}} and {{Transfer Learning}}},
  author = {Baldi, Pierre},
  year = {2012},
  pages = {37--49},
  publisher = {{JMLR Workshop and Conference Proceedings}}
}

@inproceedings{wuLearningSortFewshot2019,
  title = {Learning to {{Sort}}: {{Few-shot Spike Sorting}} with {{Adversarial Representation Learning}}},
  booktitle = {2019 41st {{Annual International Conference}} of the {{IEEE Engineering}} in {{Medicine}} and {{Biology Society}} ({{EMBC}})},
  author = {Wu, Tong and Rátkai, Anikó and Schlett, Katalin and Grand, László and Yang, Zhi},
  year = {2019},
  pages = {713--716}
}

@article{eomDeeplearnedSpikeRepresentations2021,
  title = {Deep-Learned Spike Representations and Sorting via an Ensemble of Auto-Encoders},
  author = {Eom, Junsik and Park, In Yong and Kim, Sewon and Jang, Hanbyol and Park, Sanggeon and Huh, Yeowool and Hwang, Dosik},
  year = {2021},
  journal = {Neural Networks},
  volume = {134},
  pages = {131--142}
}

@article{hayModelsNeocorticalLayer2011,
  title = {Models of {{Neocortical Layer}} 5b {{Pyramidal Cells Capturing}} a {{Wide Range}} of {{Dendritic}} and {{Perisomatic Active Properties}}},
  author = {Hay, Etay and Hill, Sean and Schürmann, Felix and Markram, Henry and Segev, Idan},
  year = {2011},
  journal = {PLOS Computational Biology},
  volume = {7},
  number = {7},
  pages = {e1002107},
  publisher = {Public Library of Science}
}

@article{hinesNEURONSimulationEnvironment1997,
  title = {The {{NEURON Simulation Environment}}},
  author = {Hines, M. L. and Carnevale, N. T.},
  year = {1997},
  journal = {Neural Computation},
  volume = {9},
  number = {6},
  pages = {1179--1209}
}

@article{hinesNeuronToolNeuroscientists2001,
  title = {Neuron: {{A Tool}} for {{Neuroscientists}}},
  author = {Hines, M. L. and Carnevale, N. T.},
  year = {2001},
  journal = {The Neuroscientist},
  volume = {7},
  number = {2},
  pages = {123--135},
  publisher = {SAGE Publications Inc STM}
}

@article{lindenLFPyToolBiophysical2014,
  title = {{{LFPy}}: A Tool for Biophysical Simulation of Extracellular Potentials Generated by Detailed Model Neurons},
  author = {Lindén, Henrik and Hagen, Espen and Leski, Szymon and Norheim, Eivind S. and Pettersen, Klas H. and Einevoll, Gaute T.},
  year = {2014},
  journal = {Frontiers in Neuroinformatics},
  volume = {7},
  publisher = {Frontiers}
}

@article{hagenViSAPyPythonTool2015,
  title = {{{ViSAPy}}: {{A Python}} Tool for Biophysics-Based Generation of Virtual Spiking Activity for Evaluation of Spike-Sorting Algorithms},
  author = {Hagen, Espen and Ness, Torbjørn V. and Khosrowshahi, Amir and Sørensen, Christina and Fyhn, Marianne and Hafting, Torkel and Franke, Felix and Einevoll, Gaute T.},
  year = {2015},
  journal = {Journal of Neuroscience Methods},
  volume = {245},
  pages = {182--204}
}

@article{hagenMultimodalModelingNeural2018,
  title = {Multimodal {{Modeling}} of {{Neural Network Activity}}: {{Computing LFP}}, {{ECoG}}, {{EEG}}, and {{MEG Signals With LFPy}} 2.0},
  author = {Hagen, Espen and Næss, Solveig and Ness, Torbjørn V. and Einevoll, Gaute T.},
  year = {2018},
  journal = {Frontiers in Neuroinformatics},
  volume = {12},
  publisher = {Frontiers}
}

@article{buccinoMEArecFastCustomizable2021,
  title = {{{MEArec}}: {{A Fast}} and {{Customizable Testbench Simulator}} for {{Ground-truth Extracellular Spiking Activity}}},
  author = {Buccino, Alessio Paolo and Einevoll, Gaute Tomas},
  year = {2021},
  journal = {Neuroinformatics},
  volume = {19},
  number = {1},
  pages = {185--204}
}

@incollection{holmesPassiveCableModeling2009,
  title = {Passive {{Cable Modeling}}},
  booktitle = {Computational {{Modeling Methods}} for {{Neuroscientists}}},
  author = {Holmes, William R.},
  year = {2009}
}

@incollection{bretteHandbookNeuralActivity2012,
  title = {Handbook of {{Neural Activity Measurement}}},
  booktitle = {Handbook of {{Neural Activity Measurement}}},
  author = {Brette, Romain and Destexhe, Alain},
  year = {2012},
  pages = {92--135},
  publisher = {Cambridge University Press},
  googlebooks = {YLyGmfVuBsIC},
  isbn = {978-0-521-51622-8}
}

@article{buzsakiOriginExtracellularFields2012,
  title = {The Origin of Extracellular Fields and Currents — {{EEG}}, {{ECoG}}, {{LFP}} and Spikes},
  author = {Buzsáki, György and Anastassiou, Costas A. and Koch, Christof},
  year = {2012},
  journal = {Nature Reviews Neuroscience},
  volume = {13},
  number = {6},
  pages = {407--420},
  publisher = {Nature Publishing Group}
}

@article{einevollModellingAnalysisLocal2013,
  title = {Modelling and Analysis of Local Field Potentials for Studying the Function of Cortical Circuits},
  author = {Einevoll, Gaute T. and Kayser, Christoph and Logothetis, Nikos K. and Panzeri, Stefano},
  year = {2013},
  journal = {Nature Reviews Neuroscience},
  volume = {14},
  number = {11},
  pages = {770--785},
  publisher = {Nature Publishing Group}
}

@inproceedings{vaswaniAttentionAllYou2017,
  title = {Attention Is All You Need},
  booktitle = {Proceedings of the 31st {{International Conference}} on {{Neural Information Processing Systems}}},
  author = {Vaswani, Ashish and Shazeer, Noam and Parmar, Niki and Uszkoreit, Jakob and Jones, Llion and Gomez, Aidan N. and Kaiser, Lukasz and Polosukhin, Illia},
  year = {2017},
  series = {{{NIPS}}'17},
  pages = {6000--6010},
  publisher = {Curran Associates Inc.},
  address = {Red Hook, NY, USA},
  isbn = {978-1-5108-6096-4}
}

@article{dorkenwaldMultilayeredMapsNeuropil2023,
  title = {Multi-Layered Maps of Neuropil with Segmentation-Guided Contrastive Learning},
  author = {Dorkenwald, Sven and Li, Peter H. and Januszewski, Michał and Berger, Daniel R. and {Maitin-Shepard}, Jeremy and Bodor, Agnes L. and Collman, Forrest and {Schneider-Mizell}, Casey M. and {da Costa}, Nuno Maçarico and Lichtman, Jeff W. and Jain, Viren},
  year = {2023},
  journal = {Nature Methods},
  volume = {20},
  number = {12},
  pages = {2011--2020},
  publisher = {Nature Publishing Group}
}

@article{yuVivoCelltypeBrain2024,
  title = {In Vivo Cell-Type and Brain Region Classification via Multimodal Contrastive Learning},
  author = {Yu, Han and Lyu, Hanrui and Xu, Ethan Yixun and Windolf, Charlie and Lee, Eric Kenji and Yang, Fan and Shelton, Andrew M. and Olsen, Shawn and Minavi, Sahar and Winter, Olivier and Laboratory, International Brain and Dyer, Eva L. and Chandrasekaran, Chandramouli and Steinmetz, Nicholas A. and Paninski, Liam and Hurwitz, Cole},
  year = {2024},
  journal = {bioRxiv Preprint},
  pages = {2024.11.05.622159},
  chapter = {New Results}
}

@inproceedings{schroffFaceNetUnifiedEmbedding2015,
  title = {{{FaceNet}}: {{A}} Unified Embedding for Face Recognition and Clustering},
  booktitle = {2015 {{IEEE Conference}} on {{Computer Vision}} and {{Pattern Recognition}} ({{CVPR}})},
  author = {Schroff, Florian and Kalenichenko, Dmitry and Philbin, James},
  year = {2015},
  pages = {815--823}
}

@article{eckartApproximationOneMatrix1936,
  title = {The Approximation of One Matrix by Another of Lower Rank},
  author = {Eckart, Carl and Young, Gale},
  year = {1936},
  journal = {Psychometrika},
  volume = {1},
  number = {3},
  pages = {211--218}
}

@article{cunninghamDimensionalityReductionLargescale2014,
  title = {Dimensionality Reduction for Large-Scale Neural Recordings},
  author = {Cunningham, John P. and Yu, Byron M.},
  year = {2014},
  journal = {Nature Neuroscience},
  volume = {17},
  number = {11},
  pages = {1500--1509},
  publisher = {Nature Publishing Group}
}

@inproceedings{choLearningPhraseRepresentations2014,
  title = {Learning {{Phrase Representations}} Using {{RNN Encoder}}–{{Decoder}} for {{Statistical Machine Translation}}},
  booktitle = {Proceedings of the 2014 {{Conference}} on {{Empirical Methods}} in {{Natural Language Processing}} ({{EMNLP}})},
  author = {Cho, Kyunghyun and Van Merrienboer, Bart and Gulcehre, Caglar and Bahdanau, Dzmitry and Bougares, Fethi and Schwenk, Holger and Bengio, Yoshua},
  year = {2014},
  pages = {1724--1734},
  publisher = {Association for Computational Linguistics},
  address = {Doha, Qatar}
}

@misc{mcinnesUMAPUniformManifold2020,
  title = {{{UMAP}}: {{Uniform Manifold Approximation}} and {{Projection}} for {{Dimension Reduction}}},
  author = {McInnes, Leland and Healy, John and Melville, James},
  year = {2020},
  number = {arXiv:1802.03426},
  eprint = {1802.03426},
  primaryclass = {stat},
  publisher = {arXiv},
  archiveprefix = {arXiv}
}

@article{dempsterMaximumLikelihoodIncomplete1977,
  title = {Maximum {{Likelihood}} from {{Incomplete Data Via}} the {{EM Algorithm}}},
  author = {Dempster, A. P. and Laird, N. M. and Rubin, D. B.},
  year = {1977},
  journal = {Journal of the Royal Statistical Society: Series B (Methodological)},
  volume = {39},
  number = {1},
  pages = {1--22}
}

@article{chengMeanShiftMode1995,
  title = {Mean Shift, Mode Seeking, and Clustering},
  author = {Cheng, Yizong},
  year = {1995},
  journal = {IEEE Transactions on Pattern Analysis and Machine Intelligence},
  volume = {17},
  number = {8},
  pages = {790--799}
}

@article{maglandSpikeForestReproducibleWebfacing2020,
  title = {{{SpikeForest}}, Reproducible Web-Facing Ground-Truth Validation of Automated Neural Spike Sorters},
  author = {Magland, Jeremy and Jun, James J and Lovero, Elizabeth and Morley, Alexander J and Hurwitz, Cole Lincoln and Buccino, Alessio Paolo and Garcia, Samuel and Barnett, Alex H},
  editor = {Meister, Markus and Calabrese, Ronald L and Meister, Markus},
  year = {2020},
  journal = {eLife},
  volume = {9},
  pages = {e55167},
  publisher = {eLife Sciences Publications, Ltd}
}

@article{hubert1985comparing,
  title={Comparing partitions},
  author={Hubert, Lawrence and Arabie, Phipps},
  journal={Journal of classification},
  volume={2},
  pages={193--218},
  year={1985},
  publisher={Springer}
}

@incollection{macqueenMethodsClassificationAnalysis1967,
  title = {Some Methods for Classification and Analysis of Multivariate Observations},
  booktitle = {Proceedings of the {{Fifth Berkeley Symposium}} on {{Mathematical Statistics}} and {{Probability}}, {{Volume}} 1: {{Statistics}}},
  author = {MacQueen, J.},
  year = {1967},
  volume = {5.1},
  pages = {281--298},
  publisher = {University of California Press}
}

@inproceedings{ngSpectralClusteringAnalysis2001,
  title = {On {{Spectral Clustering}}: {{Analysis}} and an Algorithm},
  booktitle = {Advances in {{Neural Information Processing Systems}}},
  author = {Ng, Andrew and Jordan, Michael and Weiss, Yair},
  year = {2001},
  volume = {14},
  publisher = {MIT Press}
}

@article{buccinoSpikeInterfaceUnifiedFramework2020,
  title = {{{SpikeInterface}}, a Unified Framework for Spike Sorting},
  author = {Buccino, Alessio P and Hurwitz, Cole L and Garcia, Samuel and Magland, Jeremy and Siegle, Joshua H and Hurwitz, Roger and Hennig, Matthias H},
  editor = {Colgin, Laura L and Grün, Sonja and Kloosterman, Fabian},
  year = {2020},
  journal = {eLife},
  volume = {9},
  pages = {e61834},
  publisher = {eLife Sciences Publications, Ltd}
}

@article{gouwensSystematicGenerationBiophysically2018,
  title = {Systematic Generation of Biophysically Detailed Models for Diverse Cortical Neuron Types},
  author = {Gouwens, Nathan W. and Berg, Jim and Feng, David and Sorensen, Staci A. and Zeng, Hongkui and Hawrylycz, Michael J. and Koch, Christof and Arkhipov, Anton},
  year = {2018},
  journal = {Nature Communications},
  volume = {9},
  number = {1},
  pages = {710},
  publisher = {Nature Publishing Group}
}

@article{billehSystematicIntegrationStructural2020,
  title = {Systematic {{Integration}} of {{Structural}} and {{Functional Data}} into {{Multi-scale Models}} of {{Mouse Primary Visual Cortex}}},
  author = {Billeh, Yazan N. and Cai, Binghuang and Gratiy, Sergey L. and Dai, Kael and Iyer, Ramakrishnan and Gouwens, Nathan W. and {Abbasi-Asl}, Reza and Jia, Xiaoxuan and Siegle, Joshua H. and Olsen, Shawn R. and Koch, Christof and Mihalas, Stefan and Arkhipov, Anton},
  year = 2020,
  journal = {Neuron},
  volume = {106},
  number = {3},
  pages = {388-403.e18},
  publisher = {Elsevier}
}

@article{markramReconstructionSimulationNeocortical2015,
  title = {Reconstruction and {{Simulation}} of {{Neocortical Microcircuitry}}},
  author = {Markram, Henry and Muller, Eilif and Ramaswamy, Srikanth and Reimann, Michael W. and Abdellah, Marwan and Sanchez, Carlos Aguado and Ailamaki, Anastasia and {Alonso-Nanclares}, Lidia and Antille, Nicolas and Arsever, Selim and Kahou, Guy Antoine Atenekeng and Berger, Thomas K. and Bilgili, Ahmet and Buncic, Nenad and Chalimourda, Athanassia and Chindemi, Giuseppe and Courcol, Jean-Denis and Delalondre, Fabien and Delattre, Vincent and Druckmann, Shaul and Dumusc, Raphael and Dynes, James and Eilemann, Stefan and Gal, Eyal and Gevaert, Michael Emiel and Ghobril, Jean-Pierre and Gidon, Albert and Graham, Joe W. and Gupta, Anirudh and Haenel, Valentin and Hay, Etay and Heinis, Thomas and Hernando, Juan B. and Hines, Michael and Kanari, Lida and Keller, Daniel and Kenyon, John and Khazen, Georges and Kim, Yihwa and King, James G. and Kisvarday, Zoltan and Kumbhar, Pramod and Lasserre, Sébastien and Le~Bé, Jean-Vincent and Magalhães, Bruno R. C. and {Merchán-Pérez}, Angel and Meystre, Julie and Morrice, Benjamin Roy and Muller, Jeffrey and {Muñoz-Céspedes}, Alberto and Muralidhar, Shruti and Muthurasa, Keerthan and Nachbaur, Daniel and Newton, Taylor H. and Nolte, Max and Ovcharenko, Aleksandr and Palacios, Juan and Pastor, Luis and Perin, Rodrigo and Ranjan, Rajnish and Riachi, Imad and Rodríguez, José-Rodrigo and Riquelme, Juan Luis and Rössert, Christian and Sfyrakis, Konstantinos and Shi, Ying and Shillcock, Julian C. and Silberberg, Gilad and Silva, Ricardo and Tauheed, Farhan and Telefont, Martin and {Toledo-Rodriguez}, Maria and Tränkler, Thomas and Van~Geit, Werner and Díaz, Jafet Villafranca and Walker, Richard and Wang, Yun and Zaninetta, Stefano M. and DeFelipe, Javier and Hill, Sean L. and Segev, Idan and Schürmann, Felix},
  year = 2015,
  journal = {Cell},
  volume = {163},
  number = {2},
  pages = {456--492},
  publisher = {Elsevier}
}

@misc{laquitaineSpikeSortingBiases2025,
  title = {Spike Sorting Biases and Information Loss in a Detailed Cortical Model},
  author = {Laquitaine, Steeve and Imbeni, Milo and Tharayil, Joseph and Isbister, James B. and Reimann, Michael W.},
  year = 2025,
  pages = {2024.12.04.626805},
  publisher = {bioRxiv},
  archiveprefix = {bioRxiv},
  chapter = {New Results}
}

\newpage
\appendix

\setcounter{table}{0}
\setcounter{figure}{0}
\renewcommand{\thetable}{S\arabic{table}}
\renewcommand{\thefigure}{S\arabic{figure}}

\section*{Broader Impacts}
Our work aims to advance the methodology of data processing in neuroscience experiments. It does not directly lead to any negative societal impact.

\section{Implementation Details of SimSort}
\label{appendix: A}
\subsection{Data Preparation}
\subsubsection{Train datasets}
In Sec.~\ref{subsec: Dataset Preparation}, we introduced the preparation of our simulated training dataset, divided into the continuous signal dataset and the spike waveform dataset.

The \textbf{continuous signal dataset} was constructed by concatenating multiple simulated trials (e.g., 8192 trials), each containing 6,000,000 timesteps (dt = 0.1 ms) of extracellular data. Spike labels were derived by identifying peak times of intracellular action potentials and assigning binary labels to the corresponding region. This dataset, therefore, provided labeled, continuous extracellular signals for training the detection model and evaluating the overall spike sorting performance of SimSort.

The \textbf{spike waveform dataset} was constructed by extracting waveform segments centered on labeled spike events. For each spike, a 60-timestep waveform was segmented from the extracellular signals, and up to 400 such waveforms were collected per unit. Each waveform was then paired with its corresponding unit label, creating a dataset dedicated to training and evaluating the identification model.

\subsubsection{Test datasets}
\label{appendix: test datasets}

\textbf{BBP L6 dataset (simulation)}. This dataset comprised 20 simulated trials, generated as stated in Sec.~\ref{subsec: Data Generation}, using BBP layer 6 neuron models. It provided 5 ground-truth units per recording for evaluating spike sorting performance.

\textbf{Hybrid dataset}. This dataset was composed of recordings provided by Spikeforest \cite{maglandSpikeForestReproducibleWebfacing2020}. These recordings were generated by Kilosort, with waveform templates recorded at a 5 µm electrode spacing and ~1/f background noise. It consisted of two subsets: 9 static tetrode recordings and 9 drift tetrode recordings that simulated sinusoidal probe movements. For each recording, we selected units with an SNR greater than 3 (unit IDs meeting this criterion were provided by SpikeForest to ensure a fair comparison), comprising 10–15 ground-truth units per recording. The dataset we prepared supported two modes of use. To evaluate overall spike sorting and detection performance, we used the raw data as input. For assessing identification performance, we constructed a waveform dataset by extracting 500 waveforms (60-timesteps length) per unit.

\textbf{WaveClus dataset}. This dataset consisted of four subsets (8 recordings in Easy1, 4 recordings in Easy2, 4 recordings in Difficult1, and 4 recordings in Difficult2). Each recording was generated by combining spike waveform templates derived from experimental recordings with background noise, with noise levels ranging from 0.05 to 0.4, and contained 3 ground-truth units. To increase the overall difficulty and adapt the original single-channel recording for tetrode analysis, we preprocessed the signals by duplicating them and modifying them with small random noise (standard deviation 0.01), amplitude scaling (ranging from 1 to 0.5), and random temporal shifts (-5 to +5 samples).

\textbf{IBL Neuropixels dataset}. This dataset was derived from a real-world extracellular recording, CortexLab KS046, published by the International Brain Laboratory (IBL). This recording was captured with a Neuropixels 1.0 probe across multiple brain regions from a mouse performing a decision-making behavior task. To compare spike identification performance with CEED, we prepared this dataset following the method described in the CEED paper (CEED trained and evaluated datasets derived from DanLab DY016 and DY009 Neuropixels recordings in the IBL database). We extracted waveforms from 400 units detected by Kilosort 2.5 and created fifty test sets for comparison between SimSort and CEED. Each test set contained 10 random neurons and 100 spikes per neuron. Note that CEED inputs 11-channel data into the model, while SimSort processes data from the center 4 channels.

\textbf{Real-world Neural Recording}. These recordings were collected by our wet lab (approved by research ethics committee). A 16-channel tetrode was implanted unilaterally in the primary visual cortex (V1) of mice. During the recording sessions, the mice were head-fixed, and the contralateral eye was stimulated with drifting grating stimuli of 8 different orientations. The stimuli were presented in a randomized sequence over 8 repeated trials. Each grating had a spatial frequency of 0.04 cycles per degree (cpd) and a temporal frequency of 2 Hz, lasting for 1.5 seconds per trial.

To quantify the orientation tuning of individual neurons, we computed the global orientation selectivity index (gOSI), as defined in the equation below:

\begin{equation*}
gOSI = \frac{ \left\| \sum R(\theta) e^{2i\theta} \right\| }{ \sum R(\theta) }
\end{equation*}

where $\theta$ is the direction of the moving grating and $R(\theta)$ is the neuronal response to direction $\theta$. The imaginary unit is denoted as $i$ whose square is $-1$. The gOSI ranges from 0 to 1, with 0 indicating no orientation selectivity (equal response to all directions), and 1 indicating maximal selectivity (response only to a single orientation).

\begin{table}[h]
\centering
\caption{
    \label{tab:datasets}
        Details information of test datasets used for evaluation.
    }
\vspace{1mm}
\resizebox{\linewidth}{!}{
\begin{tabular}{lcccccc}
\toprule
Datasets & \makecell{Num. \\ Recordings} & \makecell{Sample Rate \\ (Hz)} & \makecell{Num. \\ Channels} & \makecell{Duration \\ (sec per rec.)} & \makecell{Num. True Units \\ (per rec.)} & \makecell{Ground-Truth \\ Determination} \\ 
\midrule
BBP L6 dataset          & 20 & 10000 & 4 & 600         & 5        & Simulation     \\
Hybrid dataset          & 18 & 30000 & 4 & 600 / 1200 & 10-15 & Real waveforms + background noise \\
WaveClus dataset        & 20 & 24000 & 4 & 60          & 3        & Real waveforms + background noise \\
IBL Neuropixels dataset & /  & 30000 & 11 & /           & 10       & Sorted with KiloSort 2.5    \\

Real-world Neural recording & 10 & 40000 & 4 & $\approx 250$ & / & No Ground-truth \\ 
\bottomrule
\end{tabular}
}
\end{table}

\subsection{Technical Details of Simulation}

\textbf{Intracellular Simulation}. Intracellular electrophysiological activity was simulated using the NEURON package \cite{hinesNEURONSimulationEnvironment1997, hinesNeuronToolNeuroscientists2001}. Pink noise, scaled to the rheobase (the minimum current needed to elicit an action potential in a neuron), was injected into each neuron's soma to evoke stochastic action potentials, simulating biologically realistic background noise. The temporal resolution was set to 0.1 ms, and simulations spanned a total duration of 600 seconds. Neurons were initialized with a resting membrane potential of -70 mV and maintained at a physiological temperature of 34°C. Synaptic mechanisms and biophysical properties were dynamically compiled and loaded to ensure compatibility with the NEURON simulation environment.

For each neuron, a multi-compartmental model \cite{hayModelsNeocorticalLayer2011,markramReconstructionSimulationNeocortical2015,gouwensSystematicGenerationBiophysically2018,billehSystematicIntegrationStructural2020} was used to calculate transmembrane currents \cite{lindenLFPyToolBiophysical2014, hagenViSAPyPythonTool2015, hagenMultimodalModelingNeural2018, buccinoMEArecFastCustomizable2021}. The neuron was divided into compartments, each described as an equivalent electrical circuit. The dynamics of the membrane potential in each compartment $n$ were governed by Kirchhoff's current law \cite{holmesPassiveCableModeling2009}, where the sum of all currents entering or leaving a node must equal zero. Considering that extracellular potential changed much slower than ion channel dynamics, we assumed it constant. The temporal evolution of the membrane potential $V_n$ in compartment $n$ was given by
$
g_{n,n+1}(V_{n+1} - V_n) - g_{n-1,n}(V_n - V_{n-1}) 
= C_n \frac{dV_n}{dt} + \sum_j I_n^{(j)},
$
where $g_{n,n+1}$ and $g_{n-1,n}$ represented the conductances between neighboring compartments, $C_n$ was the membrane capacitance, and $\sum_j I_n^{(j)}$ accounted for ionic currents and any externally applied currents. 

These transmembrane currents served as the source for extracellular potential modeling in the next simulation stage.

\textbf{Extracellular Simulation}. Extracellular potentials were modeled using the volume conductor theory, which links transmembrane currents to extracellular potentials recorded at electrode sites \cite{bretteHandbookNeuralActivity2012, buzsakiOriginExtracellularFields2012, einevollModellingAnalysisLocal2013}. In each simulation trial, five neurons were randomly selected from the available models and positioned within a 100 × 100 × 100 $\mu$m$^3$ cubic space. These neurons consisted of different types to represent the diversity of cortical microcircuits. A virtual tetrode electrode with four recording sites was placed randomly within the same space to capture extracellular potentials.

The extracellular potential at a given electrode site $\mathbf{r}_e$ was computed as the sum of contributions from all neuronal segments within the simulation space:
$
\phi(\mathbf{r}_e, t) = \frac{1}{4\pi\sigma} \sum_{n=1}^{N} \sum_{m=1}^{M_n} \frac{I_{n,m}(t)}{|\mathbf{r}_e - \mathbf{r}_{n,m}|},
$
where $I_{n,m}(t)$ was the transmembrane current of the $m$-th compartment of the $n$-th neuron, $\mathbf{r}_{n,m}$ was its position, $\sigma$ is the extracellular medium's conductivity, and $|\mathbf{r}_e - \mathbf{r}_{n,m}|$ was the Euclidean distance between the segment and the electrode. Here, $N$ was the total number of neurons, and $M_n$ was the number of compartments in the $n$-th neuron. A reference electrode far from the source was assumed, setting $\phi = 0$ at infinity.

\subsection{Spike Detection}
\paragraph{Data Augmentation}
For each input segment $X_i \in \mathbb{R}^{T \times C}$, where $T$ represents the number of time points and $C$ denotes the number of electrode channels, a subset of channels $\mathcal{C} \subseteq \{1, 2, \dots, C\}$ was randomly selected. A probability $p \in [0, 1]$ determined whether the following augmentations were independently applied to each channel $k \in \mathcal{C}$:

\begin{equation}
    X'_{i,k} = 
    \begin{cases} 
        X_{i,k} + \epsilon, & \text{AddWithNoise}, \\
        \beta \cdot X_{i,k}, & \text{RandomAmplitudeScaling}, \\
        X_{i,k,\lfloor t \cdot n + \epsilon_t \rfloor}, & \text{RandomTimeJitter}.
    \end{cases}
\end{equation}

where $\epsilon \sim \mathcal{N}(0, \sigma^2)$ represents Gaussian noise, $\beta \sim \mathcal{U}(1 - s, 1 + s)$ is a random amplitude scaling factor, and $\epsilon_t \sim \mathcal{U}(-\delta, \delta)$ introduces small temporal shifts. The parameter $n$ is an up-sampling factor enabling finer time resolution.

\paragraph{Model-Based Detector}
In Sec.~\ref{subsec: method-Spike detection}, we introduced a model-based detector to predict spike events. These predictions were processed through the following steps:

\textbf{1. Detection with Model:}  
The model took as input a multi-channel signal \( V \in \mathbb{R}^{t \times c} \), where \( t \) was the number of time samples and \( c \) was the number of channels. This input was processed simultaneously across all channels by the detection model, which predicted a sequence of binary labels \( \hat{y}_i(t) \) for each time point:
\begin{equation}
    \hat{y}_i(t) = f_{\text{model}}(V),
\end{equation}
where \( f_{\text{model}} \) represented the detection model. The output \( \hat{y}_i(t) \) indicated the presence or absence of spike events at each time point.

\textbf{2. Adjacent Peak Merging:}  
The predicted labels \( \hat{y}_i(t) \) were refined by merging adjacent peaks to reduce false positives and consolidate overlapping detections. To identify the most relevant peaks, the channel with the maximum summed absolute signal across all time points was selected:
\begin{equation}
    c^* = \arg\max_{c \in \{1, \dots, C\}} \sum_{t=1}^T |V(t, c)|,
\end{equation}
where \( V(t, c) \) denoted the signal value at time \( t \) for channel \( c \). Using this peak channel, adjacent peaks were grouped within a window size \( \Delta t \), and only the time point with the maximum absolute signal was retained within each group:
\begin{equation} 
    t_{\text{max}} = \arg\max_{t \in [t_s, t_e]} |V(t, c^*)|,
\end{equation}
where \( [t_s, t_e] \) denoted the boundaries of a group of adjacent peaks. 

\textbf{3. Spike Peak Extraction:}  
After merging, the final spike peak positions \( \{t_k\} \) were determined as the set of time points where \( \hat{y}'_i(t) = 1 \): 
\begin{equation}
    t_k = \{t \,|\, \hat{y}'_i(t) = 1\}.
\end{equation}

\paragraph{Threshold-Based Detector}
In Sec.~\ref{sec: comparasion of detection}, we compared our model-based detector with a threshold-based detector. To ensure a fair comparison, the processing steps for the threshold-based detector were designed to align closely with those of the model-based detector. The threshold-based method processed the signal \( V \in \mathbb{R}^{t \times c} \), where \( t \) was the number of time samples and \( c \) was the number of channels. Each channel was processed independently, as follows:

\textbf{1. Local Maxima Detection}:  
For each channel \( c \), the signal \( V(t, c) \) was scanned to identify local maxima that satisfied the following conditions:
\begin{equation}
    V(t, c) > Th_{\text{single\_ch}}, \quad V(t, c) > V(t - r, c), \quad V(t, c) > V(t + r, c),
\end{equation}
where \( Th_{\text{single\_ch}} \) was the amplitude threshold, and \( r \in \{1, \dots, \text{loc\_range}\} \). These conditions ensured that the detected peaks were above the threshold and were local maxima within the range \( r \). Peaks identified by this criterion were further validated within a broader range \( \text{long\_range} \) to ensure they represented significant spike events.

\textbf{2. Adjacent Peak Merging}:  
For each channel \( c \), adjacent peaks were grouped within a window size \( \Delta t \), and only the time point with the maximum absolute signal was retained within each group:
\begin{equation} 
    t_{\text{max}} = \arg\max_{t \in [t_s, t_e]} |V(t, c)|,
\end{equation}
where \( [t_s, t_e] \) defined the boundaries of a cluster of adjacent peaks. This step ensured that closely spaced peaks were consolidated into a single spike.

\textbf{3. Spike Peak Extraction}:  
After processing all channels independently, the final detected spike positions for channel \( c \) were determined as:
\begin{equation}
    t_k = \{t \,|\, V(t, c) = t_{\text{max}}\}.
\end{equation}

\subsection{Spike Identification}
\textbf{Waveform Snippet Extraction}.  
For each detected spike at position \( t_k \), a waveform snippet \( W_k \in \mathbb{R}^{L \times c} \) was extracted from the multi-channel continuous signal \( V(t, c) \), where \( L \) was the snippet length (number of time samples) and \( c \) was the number of channels. The snippet was centered around \( t_k \) and spanned a time window determined by the pre- and post-padding values:
\begin{equation}
    W_k = V(t_k - p_{\text{pre}} : t_k + p_{\text{post}}, :),
\end{equation}
where \( V(t, c) \) represented the input signal value at time \( t \) for channel \( c \), and \( p_{\text{pre}} \) and \( p_{\text{post}} \) were the padding values (in time samples) before and after the spike time \( t_k \), respectively.

\subsection{Definitions of Metrics}
To evaluate the performance of spike sorting algorithms, we adopted standard metrics as defined in the context of spike sorting and ground-truth comparisons \cite{maglandSpikeForestReproducibleWebfacing2020, buccinoSpikeInterfaceUnifiedFramework2020}. These metrics included \textbf{accuracy}, \textbf{precision}, and \textbf{recall}, which were derived based on the matching between the sorted spike train and the ground-truth spike train.

\textbf{1. Matching Spike Events:}  
For a given sorted spike train \(k\) and ground-truth spike train \(l\), the first step was to calculate the number of matching events (\(n^{\text{match}}_{l,k}\)). A sorted spike event \(s_j^{(k)}\) was considered a match to a ground-truth event \(t_i^{(l)}\) if the absolute time difference was less than or equal to the predefined matching window \(\Delta\), which was typically set to 1 millisecond:
\begin{equation}
n^{\text{match}}_{l,k} = \#\{ i : |t_i^{(l)} - s_j^{(k)}| \leq \Delta \text{ for some } j \}.
\end{equation}

\textbf{2. Missed Events and False Positives:}  
Using the number of matching events, we computed the number of missed events (\(n^{\text{miss}}_{l,k}\)) and false positives (\(n^{\text{fp}}_{l,k}\)) as follows:
\begin{equation}
n^{\text{miss}}_{l,k} = N_l - n^{\text{match}}_{l,k}, \quad n^{\text{fp}}_{l,k} = M_k - n^{\text{match}}_{l,k},
\end{equation}
where \(N_l\) was the total number of ground-truth events in \(l\), and \(M_k\) was the total number of sorted events in \(k\).

\textbf{3. Accuracy:}  
Accuracy (\(a_{l,k}\)) balanced the contributions of both missed events and false positives and was defined as:
\begin{equation}
a_{l,k} = \frac{n^{\text{match}}_{l,k}}{n^{\text{match}}_{l,k} + n^{\text{miss}}_{l,k} + n^{\text{fp}}_{l,k}}.
\end{equation}

\textbf{4. Precision and Recall:}  
The precision (\(p_l\)) and recall (\(r_l\)) were calculated based on the best matching sorted unit \(\hat{k}_l\) for each ground-truth unit \(l\). The best match was determined as the sorted unit \(k\) with the highest accuracy:
\begin{equation}
\hat{k}_l = \arg\max_k a_{l,k}.
\end{equation}
To identify the best matching sorted unit \(\hat{k}_l\) for each ground-truth unit \(l\), the Hungarian Algorithm was applied to establish the best one-to-one correspondence between sorted and ground-truth units. Using this best match, precision and recall were defined as:
\begin{equation}
p_l = \frac{n^{\text{match}}_{l,\hat{k}_l}}{n^{\text{match}}_{l,\hat{k}_l} + n^{\text{fp}}_{l,\hat{k}_l}}, \quad
r_l = \frac{n^{\text{match}}_{l,\hat{k}_l}}{n^{\text{match}}_{l,\hat{k}_l} + n^{\text{miss}}_{l,\hat{k}_l}}.
\end{equation}

\subsection{Hyperparameter Search}
We performed a grid search to optimize the hyperparameters for the SimSort detection and identification models. The hyperparameter ranges included the number of layers, hidden size, learning rate, number of attention heads, dropout rate, and augmentation settings. The final selected values are presented in Tables \ref{tab: hyperparameter_detector} and \ref{tab: hyperparameter_extractor}. 

\subsection{Infrastructures}
\label{appendix:infra}
Our simulation dataset are generated using CPU clusters for approximately 5000 CPU hours, which can be parallelized.
Our models are trained on NVIDIA A100/V100 GPU, approximately 20 GPU hours for each time.

\begin{table}[h]
\centering
\caption{
    \label{tab: hyperparameter_detector}
    Hyperparameters of SimSort detection model.
}
\resizebox{\linewidth}{!}{%
\begin{tabular}{ccc}
\toprule
Hyperparameters    & Description                                     & Value  \\ 
\midrule
Num Layer     & Number of Transformer encoder layers            & 4      \\
Input Size    & Dimensions of the input data                    & 4      \\
Batch Size    & Number of items processed in a single operation & 128    \\
Learning Rate & Learning rate of the model                      & 5e-4 \\
Hidden Size   & Dimension of the Transformer encoder            & 256    \\
nhead         & Number of Transformer attention heads           & 4      \\
Dropout       & Dropout probability in Transformer encoder      & 0.2    \\
Weight Decay  & Regularization parameter used to prevent overfitting & 1e-5    \\
Warmup Steps  & Steps to warm up the learning rate              & 4000 \\
Num Epochs    & Number of epochs to run                         & 20     \\
Sigmoid Threshold & Probability threshold for classifying spike events  &  0.97 \\
Noise Level   & The level of noise applied to the input data    & 0.9    \\
Maximum Time Jitter   & Maximum variation in timing (in terms of upsampled sampling points) & 5          \\
Amplitude Scale Range & The range for scaling the amplitude of the input data               & {[}0, 1{]} \\
Transform Probability & The probability to perform augmentations to input data & 0.8 \\
Max Channels          & The maximum number of data channels augmentations are applied to    & 4          \\ 
\bottomrule
\end{tabular}%
}
\end{table}

\begin{table}[h]
\centering
\caption{
    \label{tab: hyperparameter_extractor}
    Hyperparameters of SimSort identification model.
}
\resizebox{\linewidth}{!}{%
\begin{tabular}{ccc}
\toprule
Hyperparameters    & Description                                                      & Value   \\
\midrule
Num Layer     & Number of GRU layers                                             & 1       \\
Input Size    & Dimensions of the input data                                     & 4       \\
Batch Size    & Number of items processed in a single operation                  & 128     \\
Learning Rate & Learning rate of the model                                       & 1e-4  \\
Hidden Size   & Latent dimention size of GRU                                     & 512     \\
Weight Decay  & Regularization parameter used to prevent overfitting             & 1e-5 \\
Num Epochs    & Number of epochs to run                                          & 10      \\
$k$ components & Number of components to reconstruct waveform using SVD          & 5  \\
Noise Level   & The level of noise applied to the input data                     & 4       \\
Maximum Time Jitter   & Maximum variation in timing (in terms of upsampled sampling points) & 50           \\
Amplitude Scale Range & The range for scaling the amplitude of the input data               & {[}0.5, 1{]} \\
Transform Probability & The probability to perform augmentations to input data & 0.5 \\
Max Channels  & The maximum number of data channels augmentations are applied to & 4       \\
\bottomrule

\end{tabular}%
}
\end{table}

\clearpage
\section{Supplementary Figures}
\begin{figure*}[h]
    \centering
    \includegraphics[width=\linewidth]{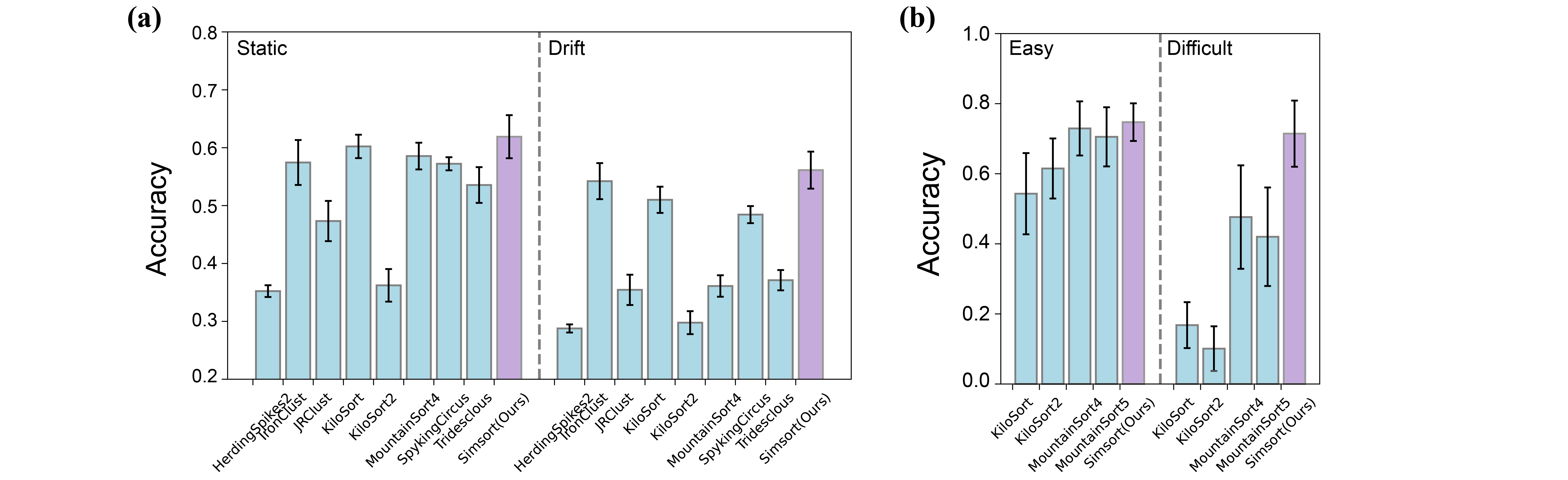}
    \caption{
    \label{fig: hybrid_waveclus-acc}
        (a) Accuracy of spike sorting on the Hybrid dataset.
        SimSort significantly outperforms HerdingSpikes2 ($p = 0.0000^*$), JRClust ($p = 0.0000^*$), KiloSort2 ($p = 0.0000^*$), and Tridesclous ($p = 0.0001^*$) in static recordings, and also shows a significant advantage over IronClust ($p = 0.0296^*$). There is no significant difference compared to KiloSort ($p = 0.5023$), MountainSort4 ($p = 0.1799$), and SpykingCircus ($p = 0.1356$). In the drift condition, SimSort significantly outperforms HerdingSpikes2 ($p = 0.0000^*$), JRClust ($p = 0.0000^*$), KiloSort ($p = 0.0285^*$), KiloSort2 ($p = 0.0000^*$), MountainSort4 ($p = 0.0004^*$), SpykingCircus ($p = 0.0111^*$), and Tridesclous ($p = 0.0007^*$), with no significant difference compared to IronClust ($p = 0.3398$). (b) Accuracy of spike sorting on the WaveClus dataset. In the easy subset, SimSort shows no significant improvement compared to KiloSort ($p = 0.0770$), KiloSort2 ($p = 0.1569$), MountainSort4 ($p = 0.6137$), and MountainSort5 ($p = 0.4424$). In the difficult subset, SimSort significantly outperforms KiloSort ($p = 0.0007^*$), KiloSort2 ($p = 0.0004^*$), and MountainSort5 ($p = 0.0449^*$), with no significant difference compared to MountainSort4 ($p = 0.0935$). Statistical analysis was performed using a paired two-tailed $t$-test.
    }
    
\end{figure*}

\begin{figure*}[h]
    \centering
    \includegraphics[width=0.5\linewidth]{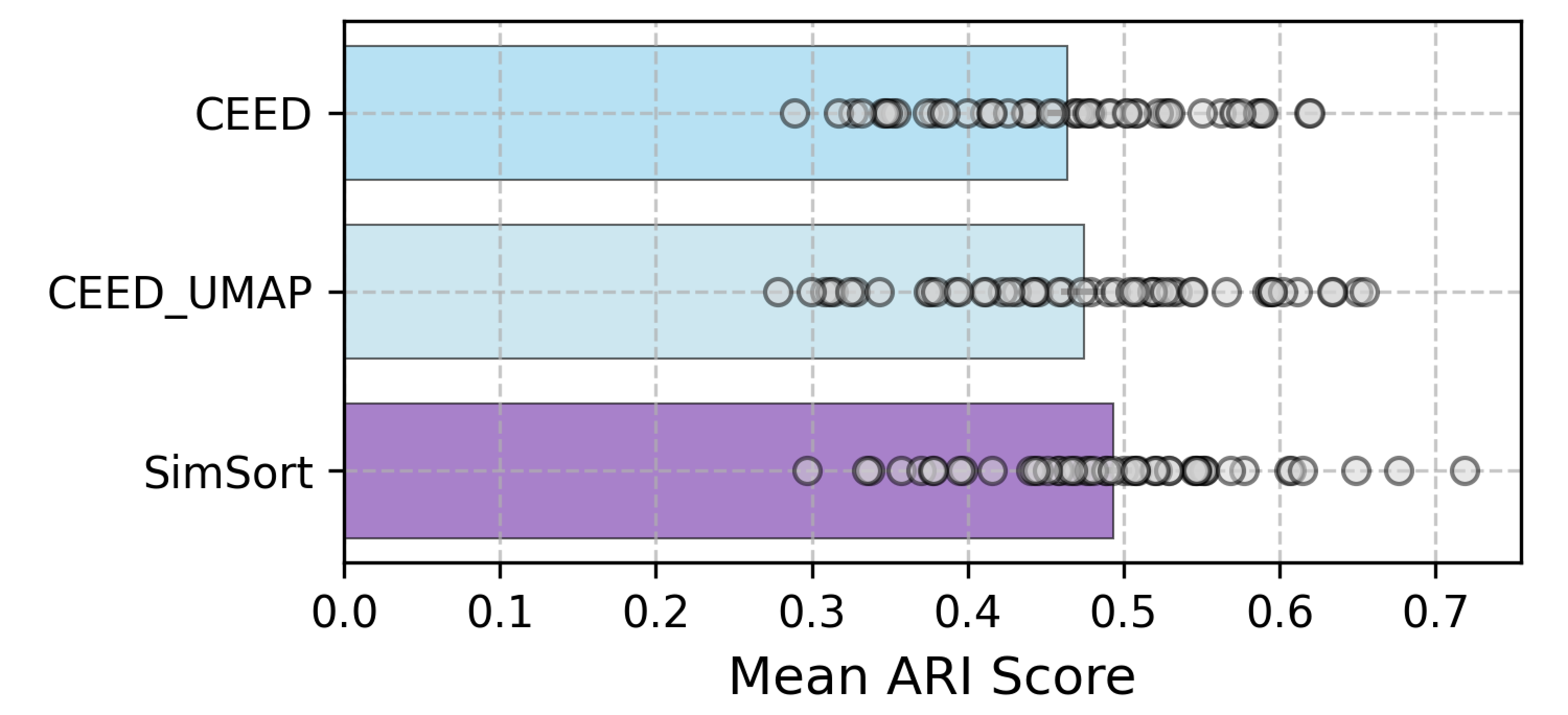}
    \caption{
    \label{fig: IBL_bar}
    Performance comparison on Fifty test sets of IBL Neuropixels dataset.
    }
\end{figure*}

\begin{table}[h]
    \centering
    \caption{
    \label{tab:IBL_identification_table}
        Performance comparison of spike identification on 50 test sets from IBL Neuropixels dataset. ARI values are presented as mean $\pm$ standard deviation. Statistical significance was assessed with paired t-tests: $p = 0.0012$ for SimSort vs. CEED, $p = 0.0554$ for SimSort vs. CEED+UMAP.
        }
    \vspace{3mm}
    \resizebox{0.5\linewidth}{!}{
    \begin{tabular}{lc}
    \toprule
    \textbf{Methods}   & \textbf{Test sets (seed 0-49)} \\
    \midrule
    CEED     & 0.46 $\pm$ 0.09 \\
    CEED+UMAP   & 0.47 $\pm$ 0.10 \\
    \rowcolor{simsortcolor!20} Ours & 0.49 $\pm$ 0.09 \\
    \bottomrule                                   
    \end{tabular}
    }
\end{table}

\begin{figure*}[h]
    \centering
    \includegraphics[height=0.9\textheight]{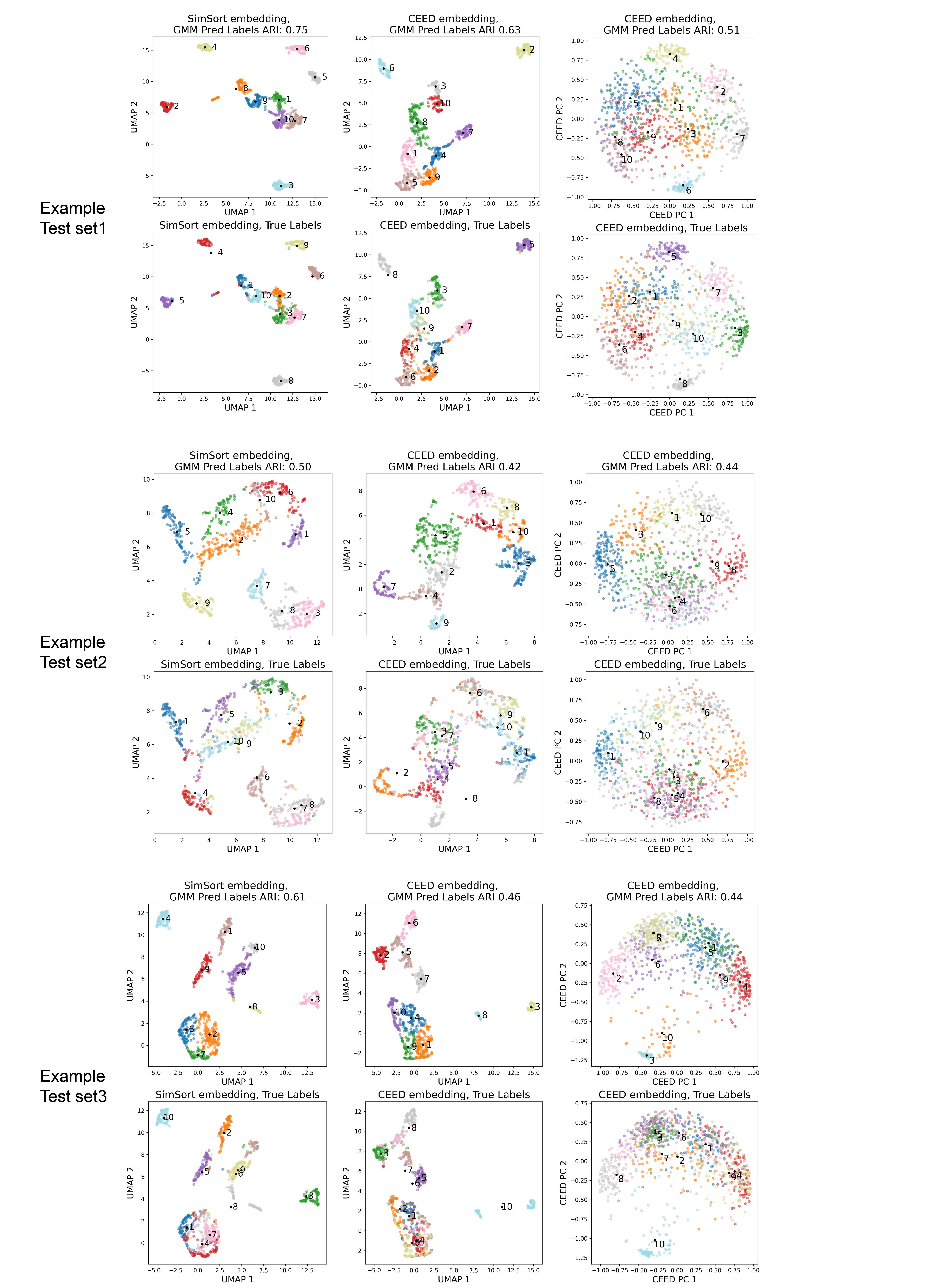}
    \caption{
    \label{fig: IBL_comparisions}
    Visualized comparison results of three example test sets from IBL Neuropixels dataset. For each example, the first column presents GMM clustering results on UMAP-embedded SimSort representations, with predicted labels (top) and true labels (bottom). The second column displays GMM clustering on UMAP-embedded CEED representations. The third column illustrates GMM clustering directly on CEED representations (as in the CEED paper).
    }
\end{figure*}

\begin{figure*}[h]
    \centering
    \vspace{-3mm}
    \includegraphics[width=0.95\linewidth]{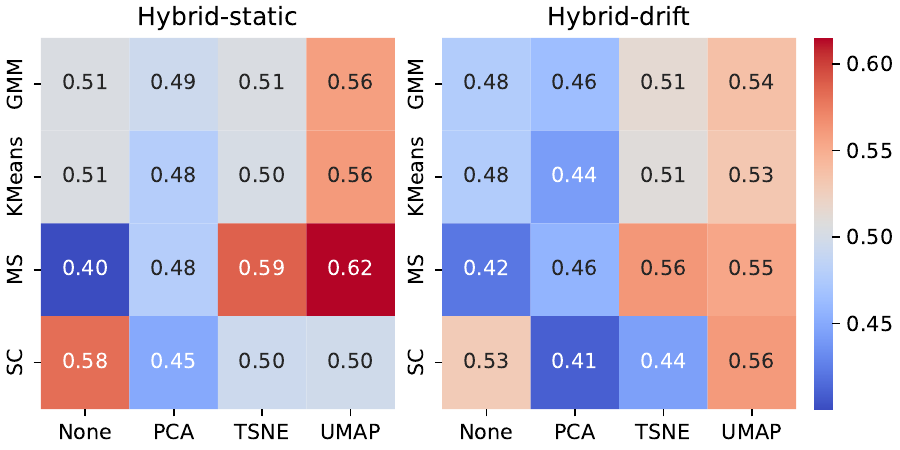}
    \caption{
    \label{fig: ablation_study}
    Impact of dimensionality reduction and clustering algorithms on the hybrid dataset. MS denotes Mean Shift, SC denotes Spectral Clustering. Comparing different combinations of dimensionality reduction methods (None, PCA, $t$-SNE, UMAP) and clustering algorithms (GMM, KMeans, Mean Shift, and Spectral Clustering) on static (left) and drift (right) subset. Here, ``None'' indicates that raw features learned by SimSort were used directly without dimensionality reduction.
    }
    \vspace{-3mm}
\end{figure*}

\clearpage
\section{More Showcases}
\label{appendix: more showcases}
In this section, we present additional visualizations from various datasets to provide a more comprehensive evaluation of SimSort's performance. These visualizations include examples of spike detection, identification, and sorting tasks. The supplementary showcases illustrate key aspects such as the alignment between detected and ground-truth spikes and the clustering of spike waveforms.

\begin{figure}[h]
    \centering
    \includegraphics[width=0.9\textwidth]{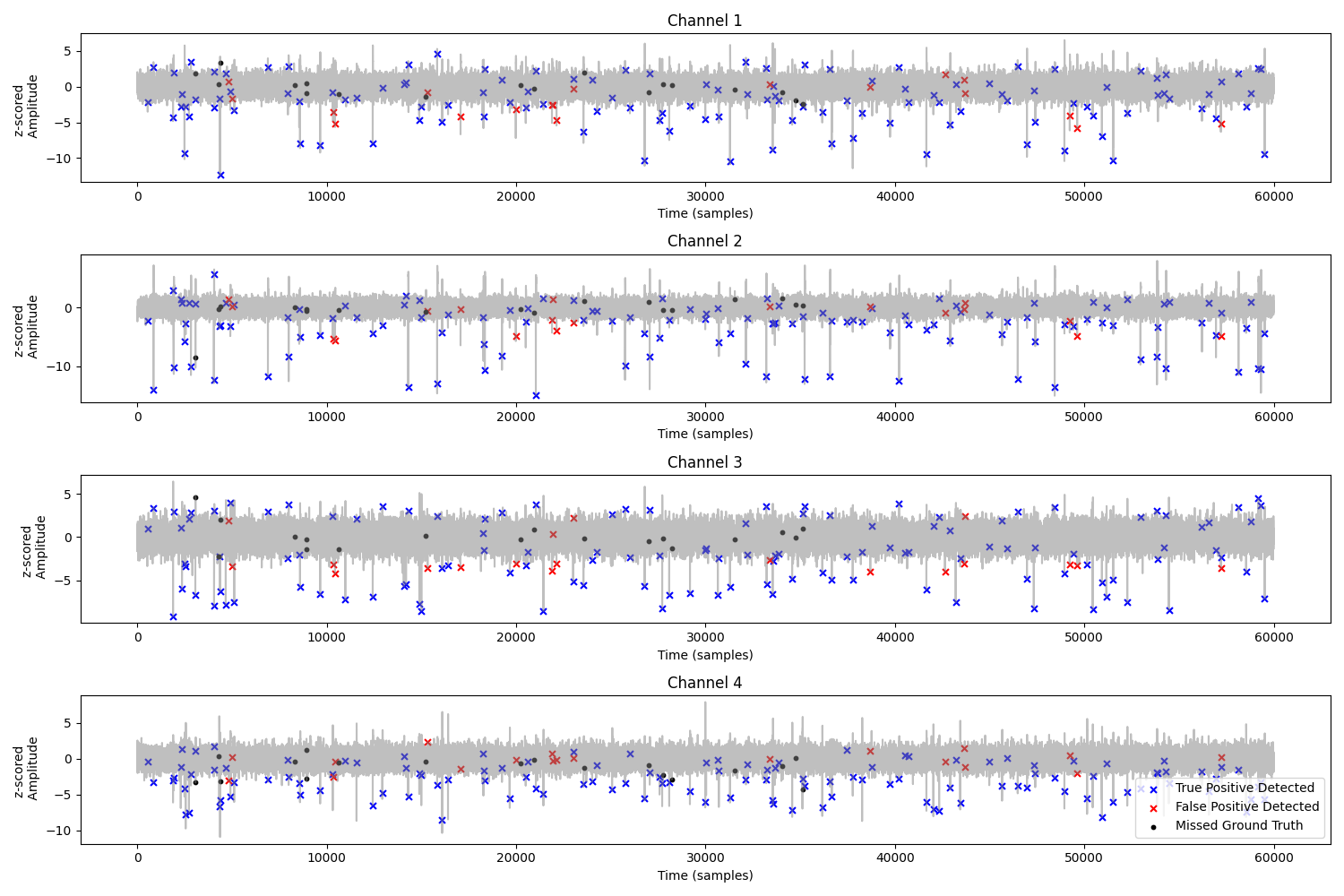}
    \caption{
    \label{fig: hybrid_detection_visualization}
    Visualization of spike detection performance on an example recording in Hybrid dataset.
    }
\end{figure}

\begin{figure}[h]
    \centering
    \includegraphics[width=0.9\textwidth]{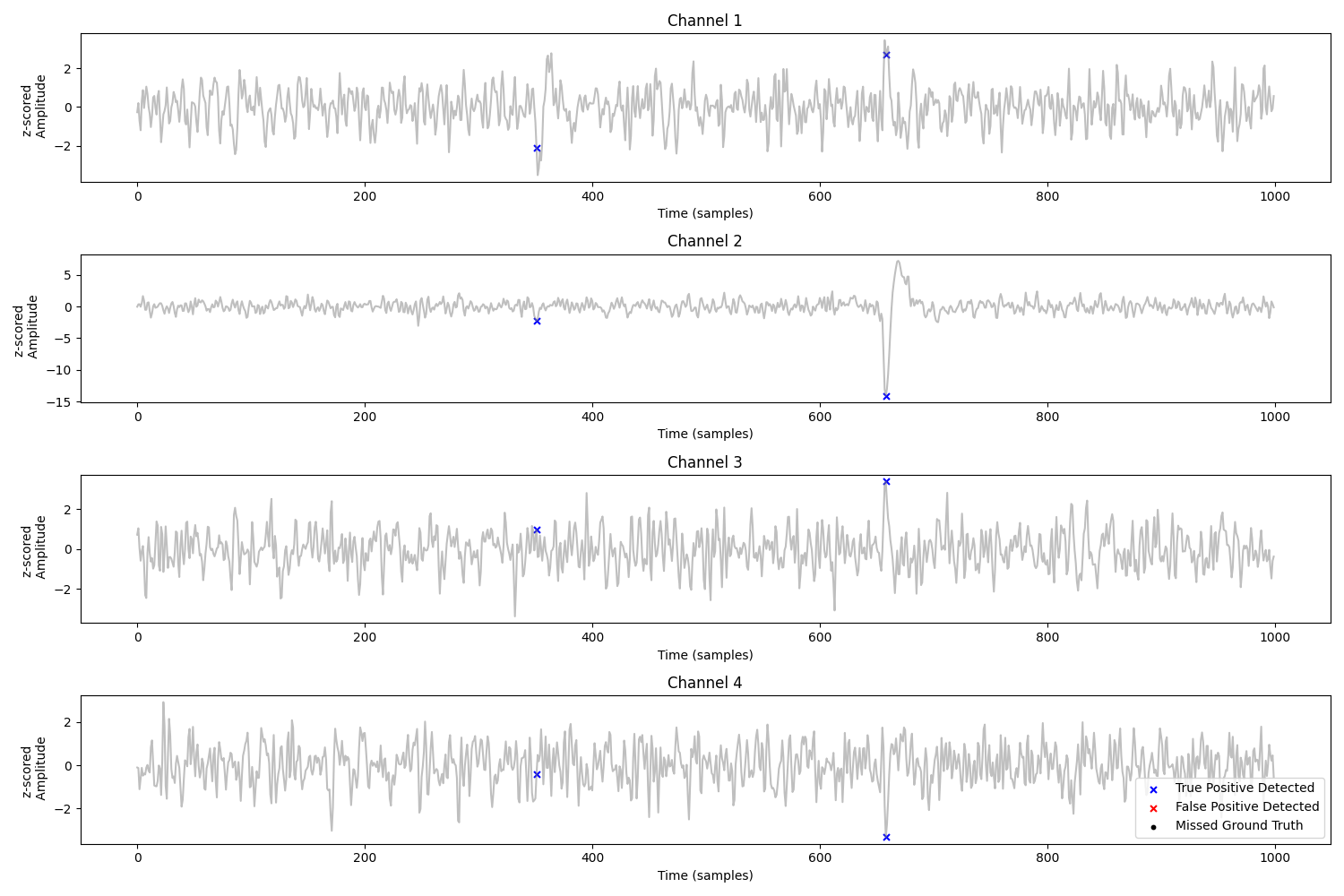}
    \caption{
    \label{fig: hybrid_detection_visualization_zoom}
    Zoomed-in visualization of spike detection performance on an example recording in Hybrid dataset.
    }
\end{figure}

\begin{figure}[h]
    \centering
    \includegraphics[width=0.9\textwidth]{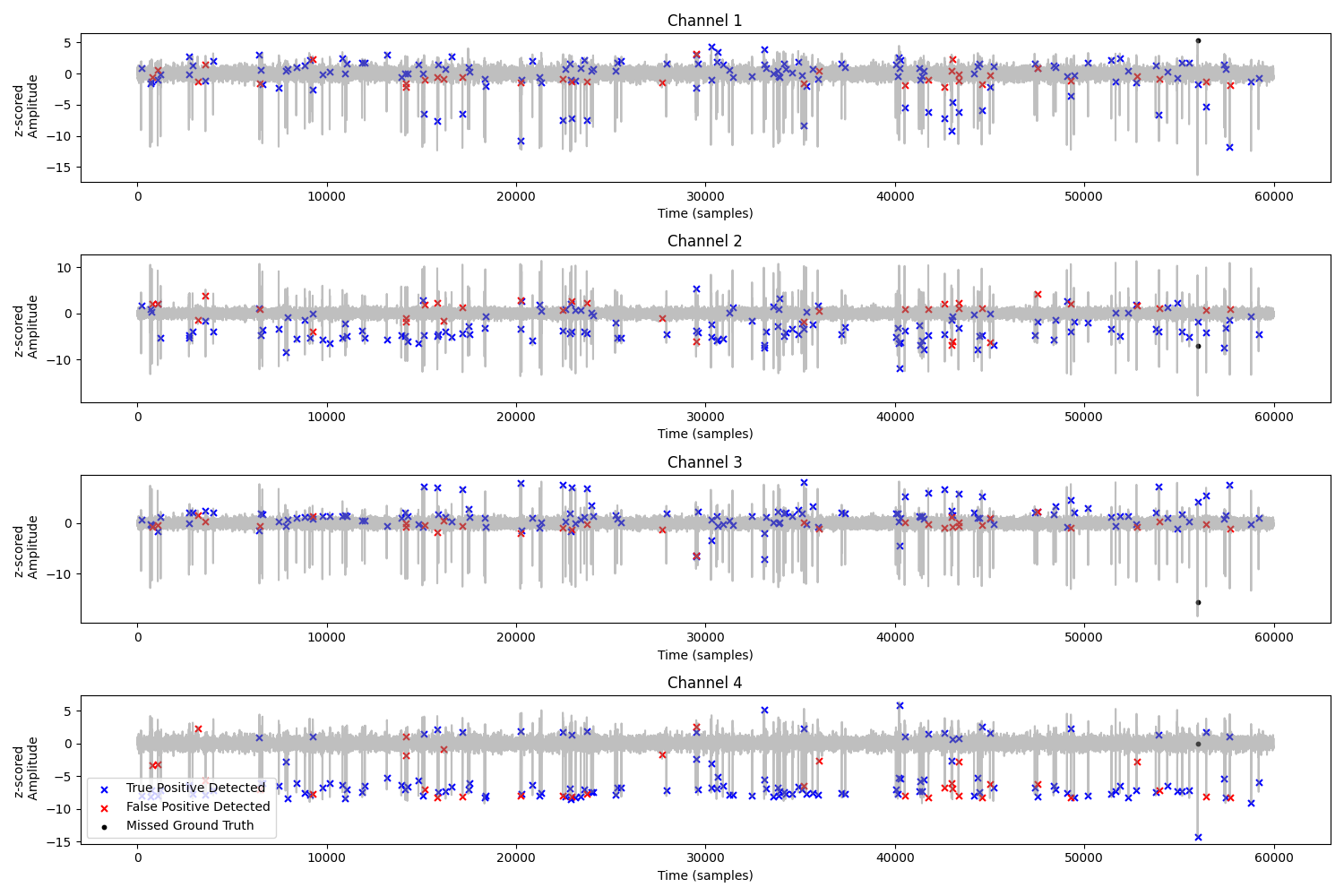}
    \caption{
    \label{fig: waveclus_detection_visualization}
    Visualization of spike detection performance on an example recording in WaveClus dataset.
    }
\end{figure}

\begin{figure}[h]
    \centering
    \includegraphics[width=0.9\textwidth]{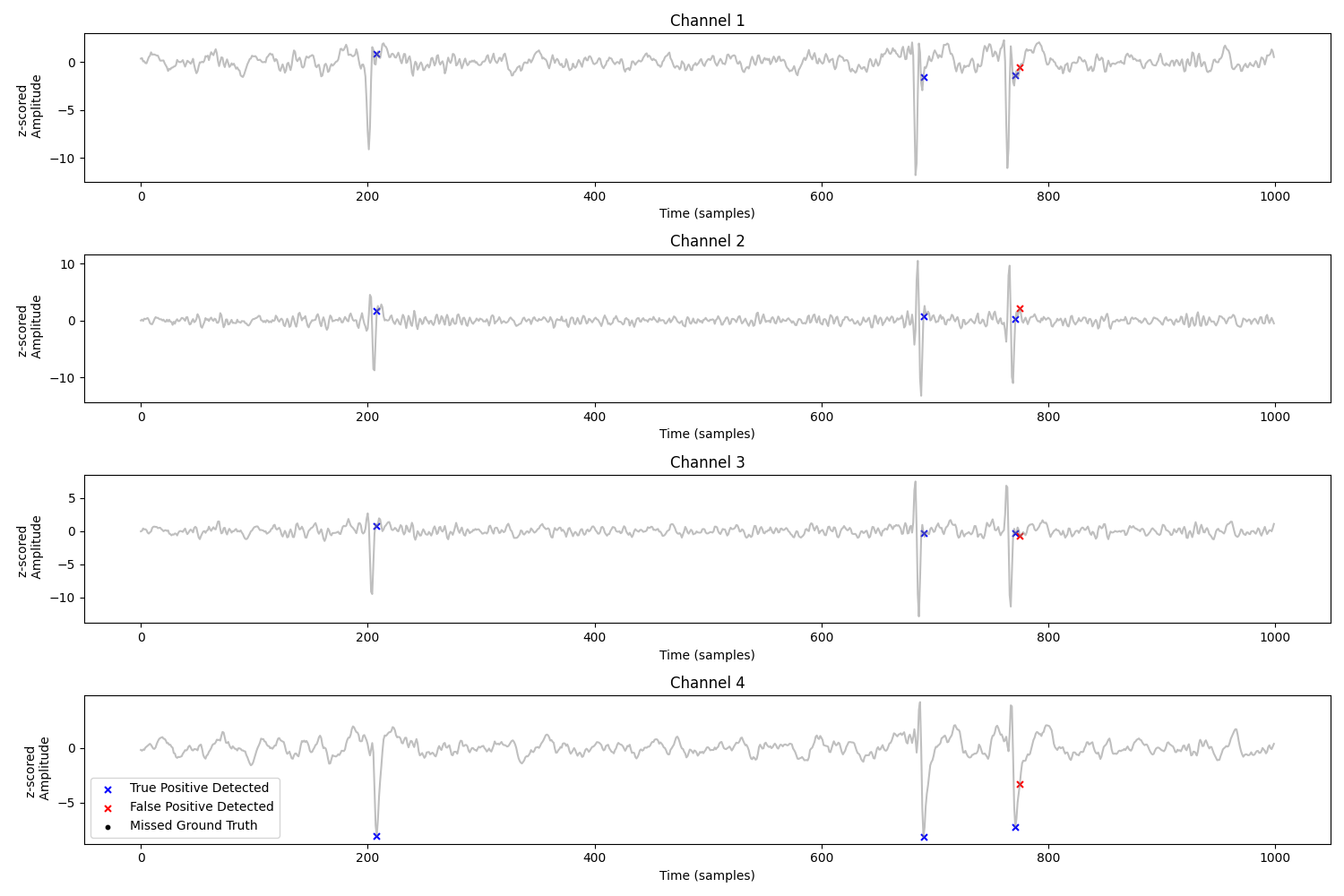}
    \caption{
    \label{fig: waveclus_detection_visualization_zoom}
    Zoomed-in visualization of spike detection performance on an example recording in WaveClus dataset.
    }
\end{figure}

\begin{figure}[h]
    \centering
    \includegraphics[width=0.9\textwidth]{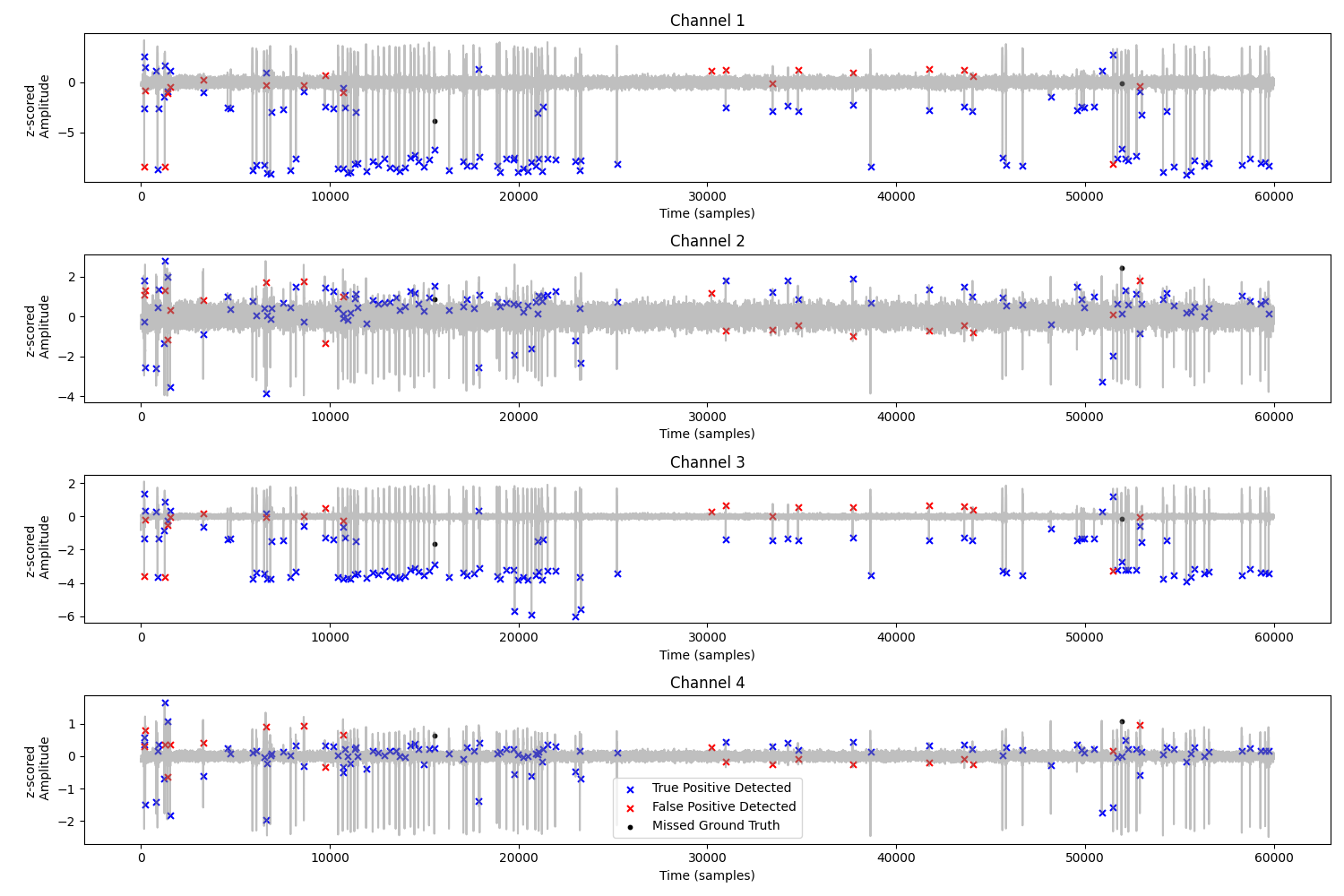}
    \caption{
    \label{fig: bbp_detection_visualization}
    Visualization of spike detection performance on an example recording in BBP L6 dataset.
    }
\end{figure}

\begin{figure}[h]
    \centering
    \includegraphics[width=0.9\textwidth]{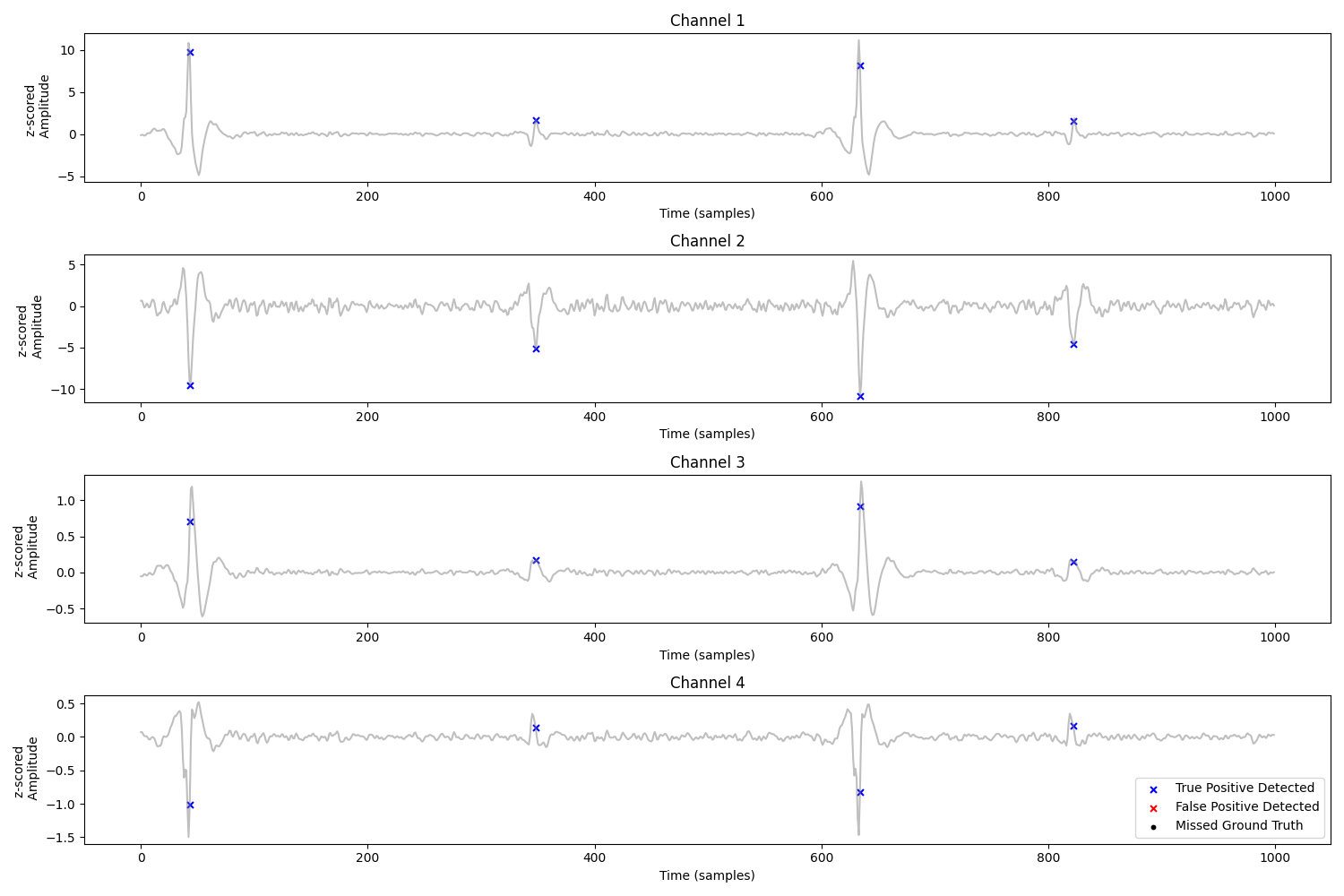}
    \caption{
    \label{fig: bbp_detection_visualization_zoom}
    Zoomed-in visualization of spike detection performance on an example recording in BBP L6 dataset.
    }
\end{figure}

\begin{figure}[h]
    \centering
    \includegraphics[height=0.9\textheight]{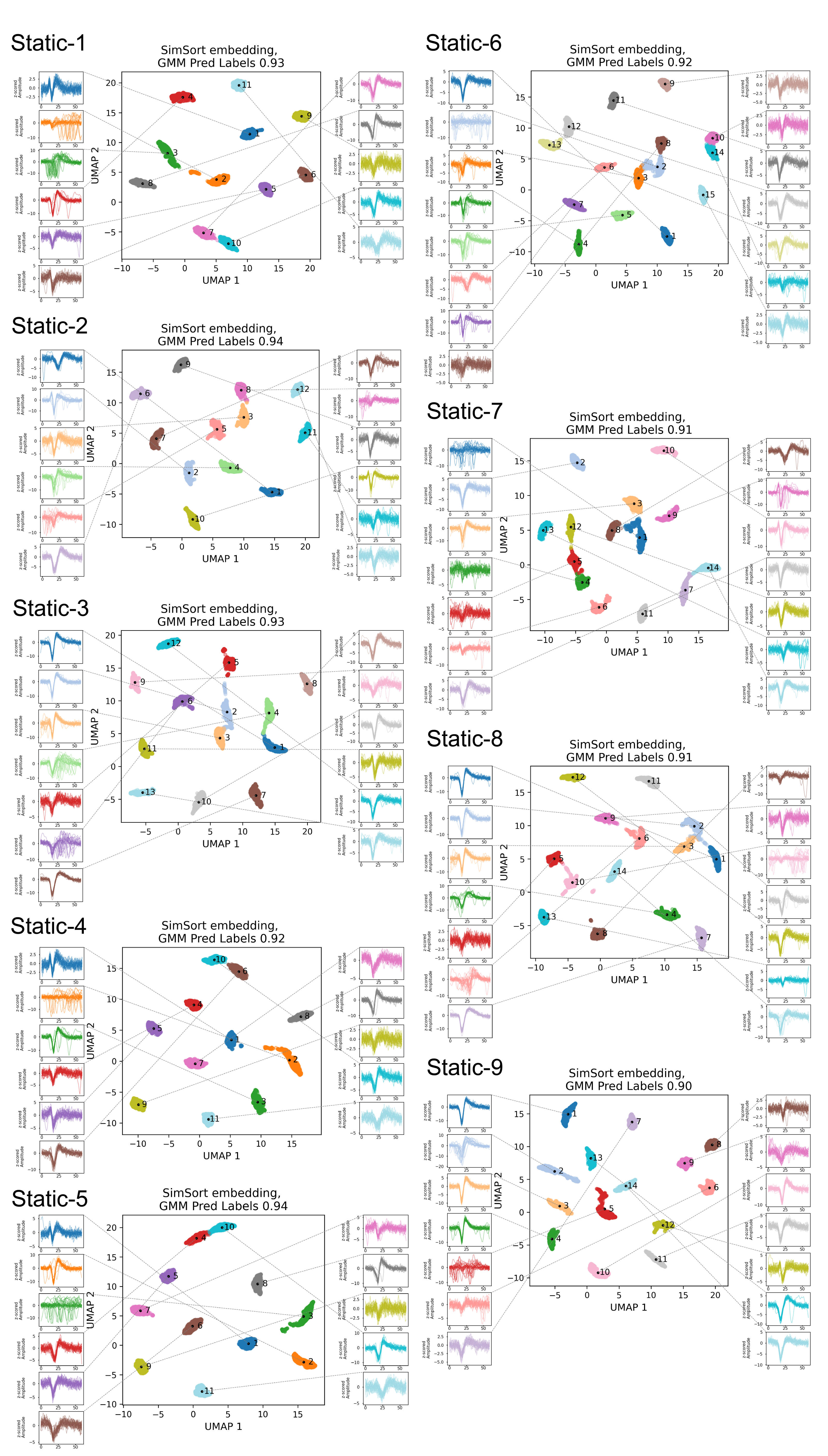}
    \caption{
    \label{fig: hybrid_static_identification_visualization_1}
    Visualization of spike identification performance on hybrid static subset.
    }
\end{figure}

\begin{figure}[h]
    \centering
    \includegraphics[height=0.9\textheight]{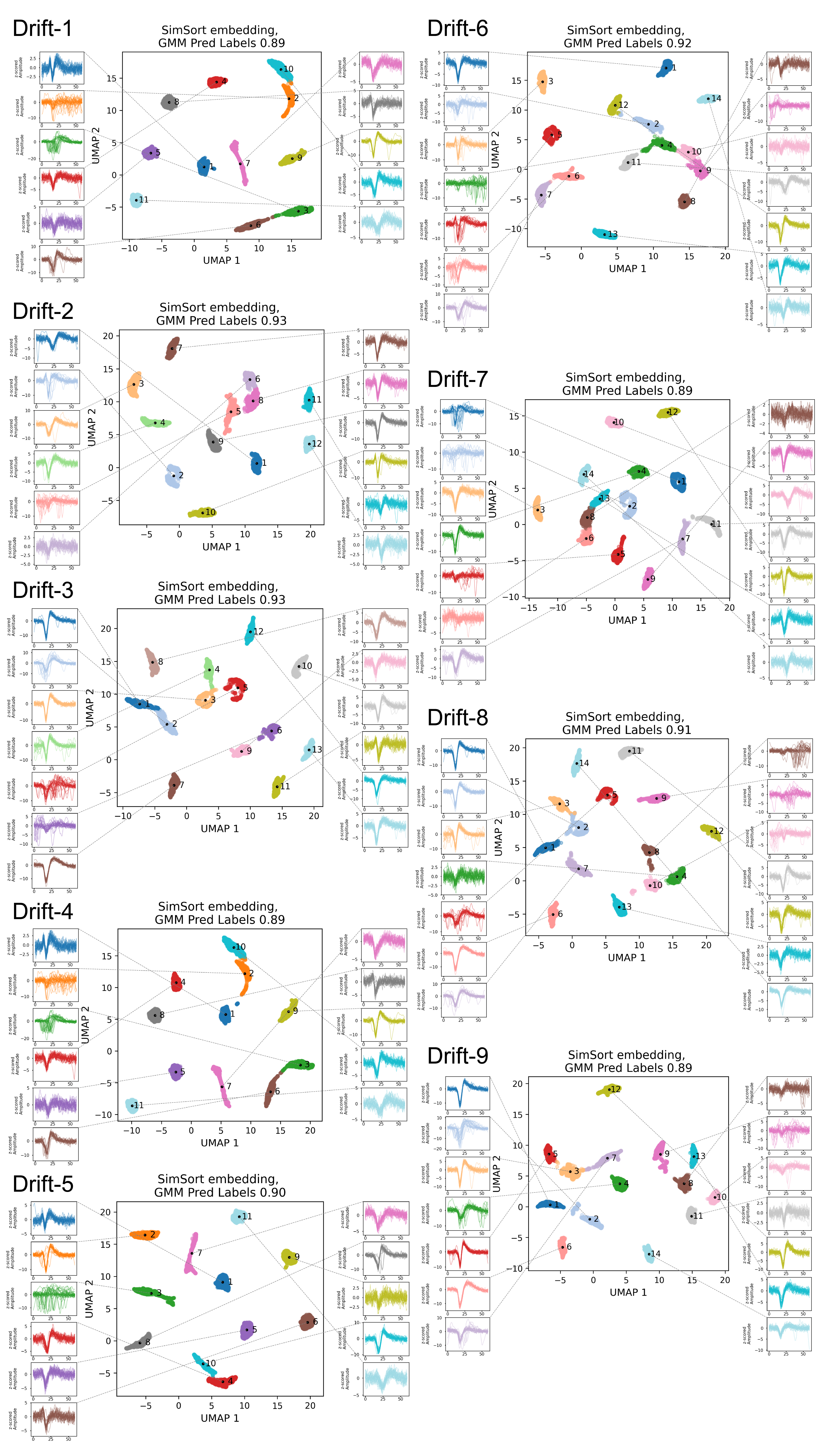}
    \caption{
    \label{fig: hybrid_drift_identification_visualization_2}
    Visualization of spike identification performance on hybrid drift subset.
    }
\end{figure}

\begin{figure}[h]
    \centering
    \includegraphics[width=\textwidth]{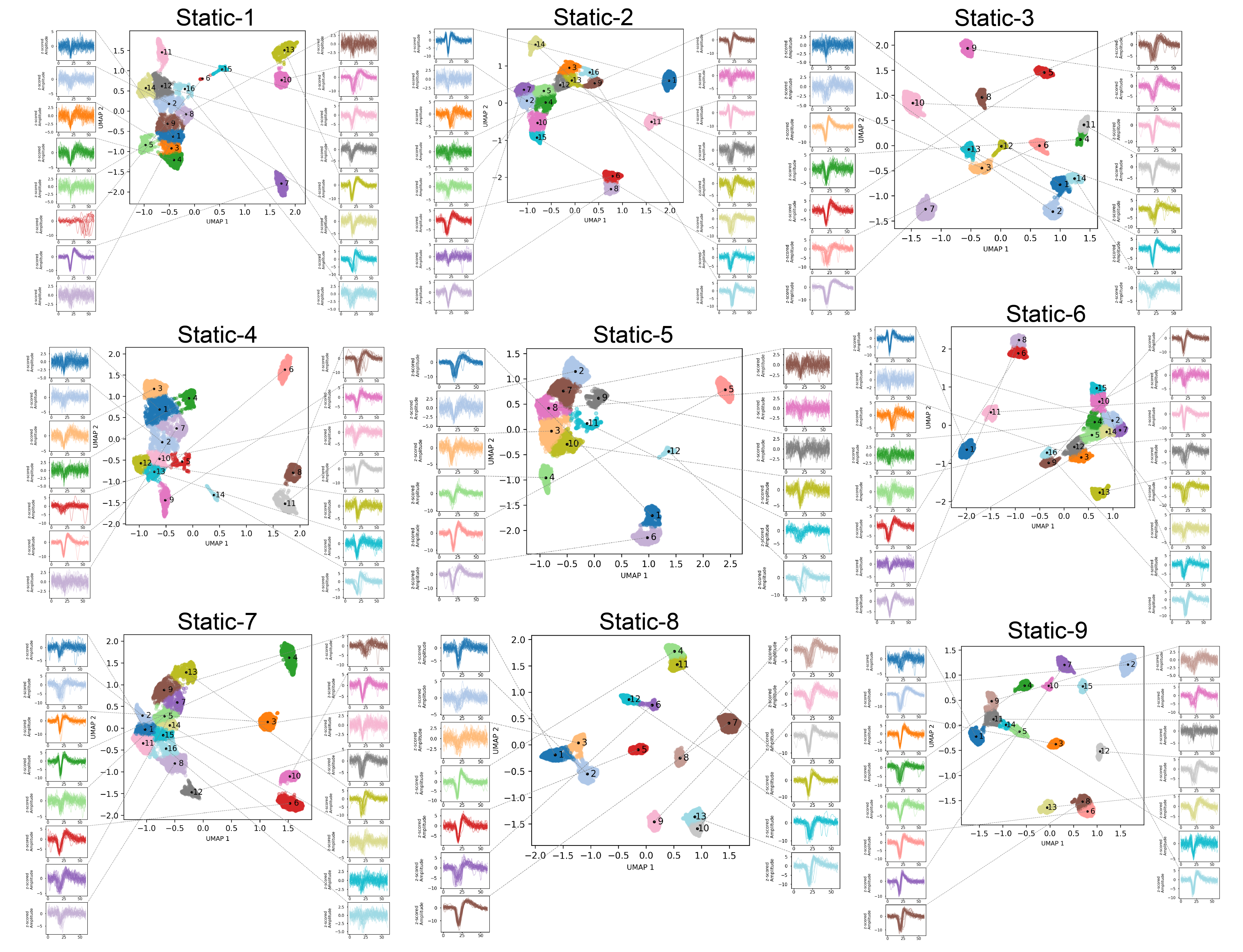}
    \caption{
    \label{fig: hybrid_static_sorting_visualization_1}
    Visualization of spike sorting performance on hybrid static subset.
    }
\end{figure}

\begin{figure}[h]
    \centering
    \includegraphics[width=\textwidth]{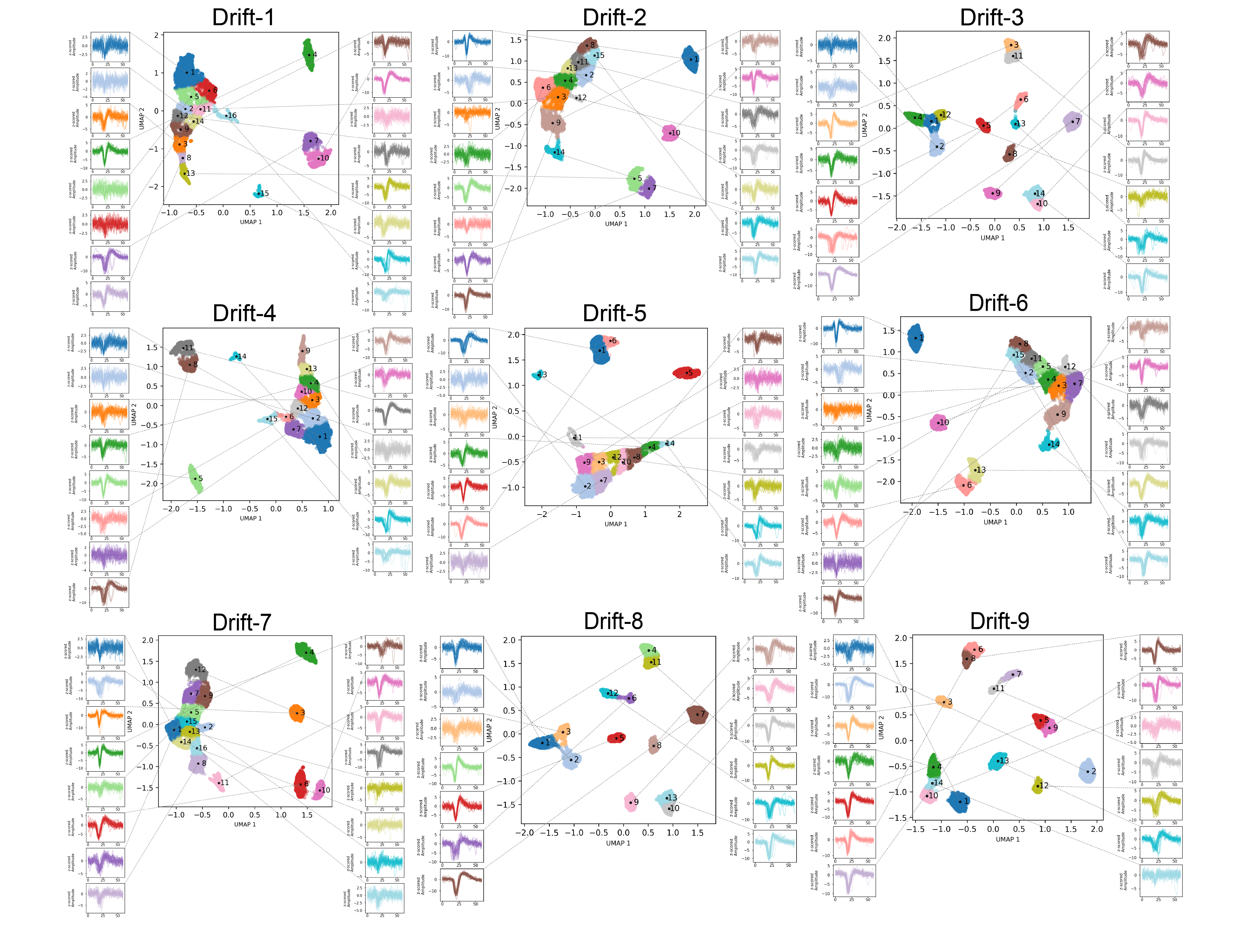}
    \caption{
    \label{fig: hybrid_drift_sorting_visualization_2}
    Visualization of spike sorting performance on hybrid drift subset.
    }
\end{figure}

\begin{figure}[h]
    \centering
    \includegraphics[width=\textwidth]{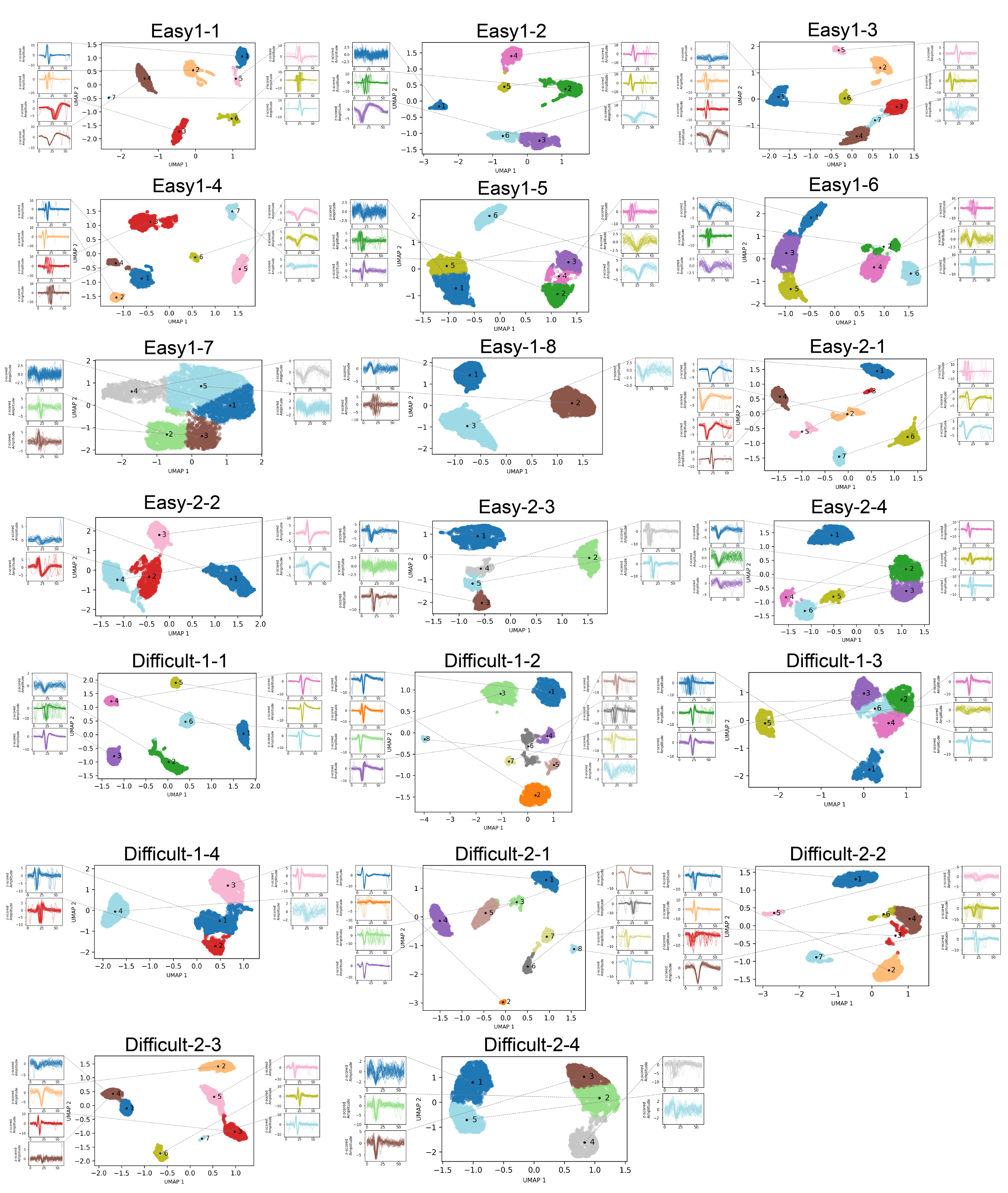}
    \caption{
    \label{fig: waveclus_sorting_visualization}
    Visualization of spike sorting performance on WaveClus dataset.
    }
    
\end{figure}

\begin{figure}[h]
    \centering
    \includegraphics[width=\textwidth]{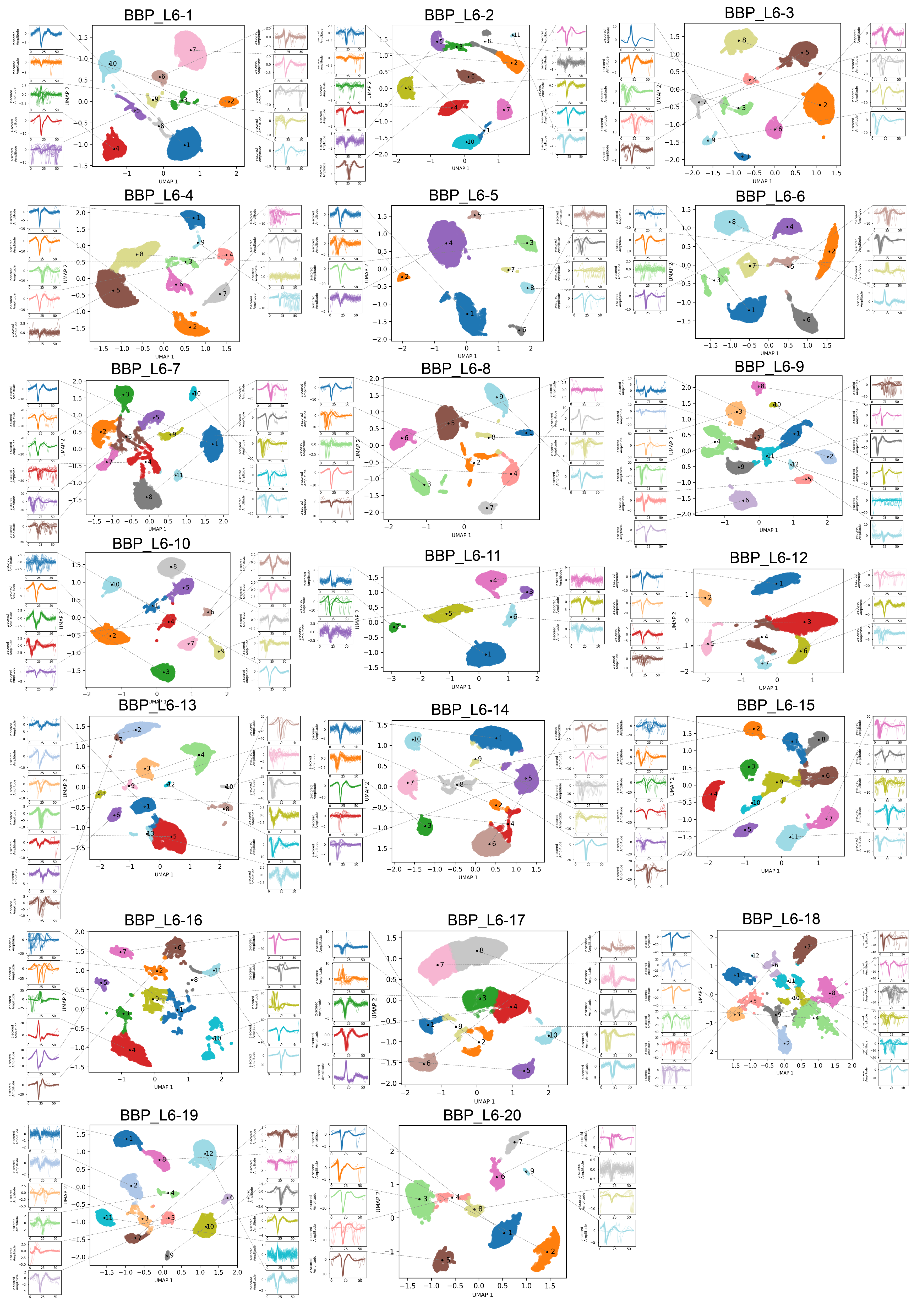}
    \caption{
    \label{fig: bbp_sorting_visualization}
    Visualization of spike sorting performance on BBP L6 dataset.
    }
\end{figure}

\begin{figure}[h]
    \centering
    \includegraphics[width=\textwidth]{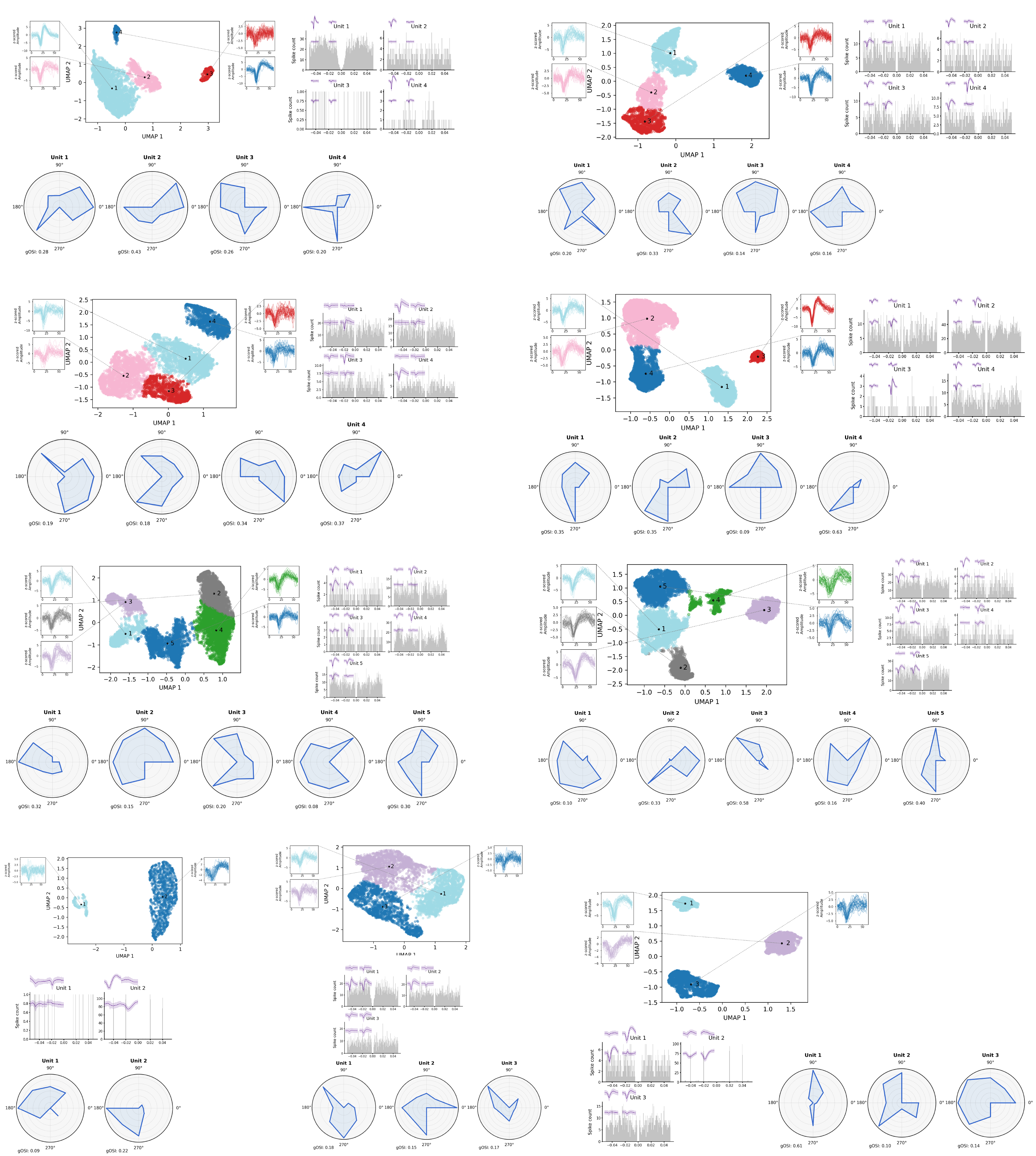}
    \caption{
    \label{fig: realdata_sorting_visualization}
    Visualization of spike sorting performance on real recording data.
    }
\end{figure}

\clearpage
\end{document}